\documentclass[10pt,conference]{IEEEtran}
\usepackage{graphicx}

\IEEEoverridecommandlockouts
 \usepackage{microtype}
\usepackage{graphicx}
\usepackage{booktabs}
\usepackage{siunitx}
\usepackage{paralist}
\usepackage{multirow}
\usepackage{array}
\usepackage{cite}
\usepackage{balance}
\usepackage{fontawesome}
\usepackage[para]{footmisc}
\usepackage{makecell}
\usepackage{hyperref}
\usepackage[hyphenbreaks]{breakurl}

\usepackage{xcolor}

\usepackage{wrapfig}
\usepackage[caption=false,font=footnotesize]{subfig}

\usepackage[T1]{fontenc}
\usepackage{float}
\usepackage[utf8]{inputenc}
\usepackage{subfiles}
\usepackage{caption}
\usepackage[noend]{algorithmic}
\usepackage{algorithm}

\usepackage{tocbasic}
\usepackage{multicol}
\usepackage[]{scrlayer-scrpage}

\usepackage{syntax}
\renewcommand{\litleft}{\sf \begin{tikz}[baseline=(X.base)]\node [draw=gray!40,fill=gray!0,semithick,rectangle,inner sep=1pt, minimum size=1em, outer sep=0pt, rounded corners=2pt] (X) }
\renewcommand{\litright}{;\end{tikz} \normalfont}

\usepackage{enumitem}
\usepackage{xspace}
\usepackage{xargs}
\usepackage{amsthm}
\usepackage{amsmath}
\usepackage{amsfonts}
\usepackage{amssymb}
\usepackage{pifont}

\newtheoremstyle{slanted}
  {0.3em}
  {0.3em}
  {\slshape}
  {}
  {\bfseries}
  {.}
  {0.5em}
  {}
 
\theoremstyle{slanted}

\newtheorem{example}{Example}

\newcommand\definetool[2]{\newcommand{#1}{{\textsc{#2}}\xspace}}
\definetool{\scratch}{Scratch}
\definetool{\whisker}{Whisker}
\definetool{\litterbox}{LitterBox}
\definetool{\bastet}{Bastet}
\definetool{\hairball}{Hairball}
\definetool{\drscratch}{Dr. Scratch}
\definetool{\qualityhound}{Quality Hound}
\definetool{\jadet}{Jadet}
\definetool{\aumextr}{GeSMo}
\definetool{\oumextractor}{OUMExtractor}
\definetool{\aums}{AUMs}
\definetool{\aum}{AUM}
\definetool{\oums}{OUMs}
\definetool{\oum}{OUM}
\definetool{\java}{Java}

\newcommand{\sm}{script model\xspace}
\newcommand{\sms}{script models\xspace}
\newtheorem{definition}{Definition}

\usepackage{multirow}
\usepackage{array}
\newlength{\defaulttabcolsep}
\definecolor{OliveGreen}{rgb}{0,0.6,0}
\usepackage[colorinlistoftodos,prependcaption,textsize=small,disable]{todonotes}
\newcommandx{\missing}[2][1=]{\todo[linecolor=red,backgroundcolor=red!25,bordercolor=red,#1]{#2}}
\newcommandx{\unsure}[2][1=]{\todo[linecolor=orange,backgroundcolor=orange!25,bordercolor=orange,#1]{#2}}
\newcommandx{\change}[2][1=]{\todo[linecolor=blue,backgroundcolor=blue!25,bordercolor=blue,#1]{#2}}
\newcommandx{\info}[2][1=]{\todo[linecolor=OliveGreen,backgroundcolor=OliveGreen!25,bordercolor=OliveGreen,#1]{#2}}

\newcommand{\done}[1]{\todo[color=green,inline]{\textbf{Done}: #1}}
\newcommand{\ignore}[1]{\todo[color=yellow,inline]{\textbf{Ignoring}: #1}}

\usepackage{fbox}
\newcommand{\summary}[2]{
        \vspace{1mm}
        \noindent
        \colorbox{gray!15}{%
            \parbox{.97\linewidth}{%
                    \textbf{#1 Summary.}
                #2
            }%
        }%
}%

\usepackage[formats]{listings}
\definecolor{lightgray}{rgb}{.9,.9,.9}
\definecolor{darkgray}{rgb}{.4,.4,.4}
\definecolor{purple}{rgb}{0.65, 0.12, 0.82}
\lstdefinelanguage{JavaScript}{
  keywords={typeof, new, true, false, catch, function, return, null, catch, switch, var, if, in, while, do, else, case, break},
  keywordstyle=\color{blue}\bfseries,
  ndkeywords={class, export, boolean, throw, implements, import, this},
  ndkeywordstyle=\color{darkgray}\bfseries,
  identifierstyle=\color{black},
  sensitive=false,
  comment=[l]{//},
  morecomment=[s]{/*}{*/},
  commentstyle=\color{purple}\ttfamily,
  stringstyle=\color{red}\ttfamily,
  morestring=[b]',
  morestring=[b]"
}

\usepackage{tikz}
\usepackage{pgflibraryshapes}
\usetikzlibrary{arrows}
\usetikzlibrary{chains}
\usetikzlibrary{fadings}
\usetikzlibrary{matrix}
\usetikzlibrary{fit}
\usetikzlibrary{positioning}
\usetikzlibrary{shapes}
\usetikzlibrary{automata}

\tikzset{
 ctrlstate/.style = {state,align=center,inner sep=2pt, minimum size=2mm},
 cfastate/.style = {ctrlstate},
 cfatargetstate/.style = {cfastate, double},
 compstate/.style = {cfastate, rounded rectangle, minimum height=5mm, minimum width=3em, inner sep=3pt},
 concept/.style = {cfastate, inner sep=3pt, fill=gray!50, rectangle, 
        minimum width=17mm, minimum height=9mm, draw=gray!6 },
 crosscutting/.style = {concept, rounded rectangle, fill=gray!30},
 conceptstate/.style = {concept, rounded rectangle, fill=gray!30, draw=gray!90, minimum width=20mm},
 inputstate/.style = {concept, rectangle, fill=gray!10, draw=gray!90, minimum width=20mm, inner sep=3pt},
 explain/.style = {circle, draw=gray!30, line width=1mm, minimum size=7mm},
 one/.style = {fill=blue!70,draw=blue!70},
 two/.style = {fill=red!50,draw=red!50},
 abststate/.style = {rectangle,align=center,inner sep=2pt,minimum size=3.5mm,fill=gray!0, draw=gray!90},
 line/.style = {draw},
 trans/.style = {draw,semithick,->,shorten >=1pt,>=stealth'},
 missing/.style = {draw=red,densely dotted,fill=red,semithick,->,shorten >=1pt,>=stealth'},
 ctrans/.style = {draw,very thick,->,shorten >=1pt,>=stealth',draw=gray!90},
 epsilon/.style = {trans,dashed},
 strengthen/.style = {draw=gray!30,semithick,double,shorten >=1pt,>=stealth',line width=1mm},
}

\newcommand{\toolname}[1]{\textsc{#1}\xspace}%
\newcommand{\blockleft}{\begin{mbox}\sf\begin{tikz}[baseline=(X.base)]\node[draw=black!60,fill=black!3,semithick,rectangle,inner sep=1pt, minimum size=1em, outer sep=0pt, rounded corners=1pt] (X)}%
\newcommand{\blockright}{;\end{tikz}\normalfont\end{mbox}}%
\newcommand{\scratchblock}[1]{\blockleft{\small\textsf{#1}}\blockright}

\newcommand{\mblockleft}{\begin{mbox}\sf\begin{tikz}[baseline=(X.base)]\node[draw=red!60,densely dotted,fill=red!3,semithick,rectangle,inner sep=1pt, minimum size=1em, outer sep=0pt, rounded corners=1pt] (X)}%
\newcommand{\mblockright}{;\end{tikz}\normalfont\end{mbox}}%
\newcommand{\mscratchblock}[1]{\mblockleft{\small\color{red}\textsf{#1}}\mblockright}

\newcommand{\cat}{\emph{Cat}\xspace}%
\newcommand{\horse}{\emph{Horse}\xspace}%
\newcommand{\elephant}{\emph{Elephant}\xspace}%
\newcommand{\monkey}{\emph{Monkey}\xspace}%
\newcommand{\fruit}{\emph{Fruit}\xspace}%
\newcommand{\open}{\emph{Open}\xspace}%

\begin{document}

\title{Finding Anomalies in Scratch Programming Assignments}
\title{Anomalies in Scratch Programming Assignments}
\title{Anomaly Detection in Scratch Programming Assignments}
\title{Anomaly Detection in Scratch Assignments}
\title{Finding Anomalies in Scratch Assignments}

\author{\IEEEauthorblockN{Nina Körber}
\IEEEauthorblockA{
\textit{University of Passau}\\
Passau, Germany}
\and
\IEEEauthorblockN{Katharina Geldreich}
\IEEEauthorblockA{
\textit{Technical University of Munich}\\
Munich, Germany}
\and
\IEEEauthorblockN{Andreas Stahlbauer}
\IEEEauthorblockA{
\textit{University of Passau}\\
Passau, Germany}
\and
\IEEEauthorblockN{Gordon Fraser}
\IEEEauthorblockA{
\textit{University of Passau}\\
Passau, Germany}
}

\maketitle

\begin{abstract}
%
%
In programming education, teachers need to monitor and assess 
the progress of their students by investigating the code they write. 
%
%
Code quality of programs written in traditional programming languages 
can be automatically assessed with automated tests, verification tools, or linters.
In many cases these approaches rely on some form of manually written
formal specification to analyze the given programs.
Writing such specifications, however, is hard for teachers,
who are often not adequately trained for this task.
Furthermore, automated tool support for popular block-based introductory programming 
languages like \scratch is lacking. 
%
%
Anomaly detection is an approach to automatically identify deviations 
of common behavior in datasets without any need for writing a specification.
In this paper, we use anomaly detection to automatically find deviations of 
\scratch code in a classroom setting, where anomalies can represent 
erroneous code, alternative solutions, or distinguished work.
%
%
%
Evaluation on solutions of different programming tasks
demonstrates that anomaly detection can successfully be applied 
to tightly specified as well as open-ended programming tasks.
\end{abstract}


\section{Introduction}\label{sec:introduction}

Teachers frequently have to evaluate students' implementations of
programming assignments to provide feedback and support, assess progress,
identify recurring problems, and to derive grades. These tasks are
challenging because they require comprehending, analyzing, and debugging different
program variants, often containing creative and unique bugs.

These tasks can be supported with automated software analysis tools;
for example, a common way to assess the correctness of student
solutions is to run automated tests.
However, programming is increasingly taught at earlier ages, often as
early as elementary school, using educational programming languages
such as \scratch. This causes several issues: First, automated tools
that are common for advanced, text-based programming languages are
rarely available for these educational programming languages. Even
when they are, teachers at elementary school level often have no
training in how to formalize specifications or automated tests; indeed
even professional developers often fail to produce adequate
tests. Finally, even a thorough test suite may fail to reveal programs
that produce the correct result using an incorrect solution path.

To address this problem, we propose the use of \emph{anomaly
  detection} for classroom programming scenarios.
Anomaly detection is based on the idea that common behavior is more
likely correct behavior, and that rare deviations of common
behavior~(so called \emph{anomalies}) are likely wrong.  In the
context of software engineering, anomaly detection has been
successfully applied to find bugs in large code bases requiring no
specification, no tests, and no manual labor. 
While code bases in an educational setting
tend to be small, they do contain 
 common code constructs
which can be exploited to find anomalies that deviate from the common
solutions.

\begin{figure}[t]
    \centering
    \subfloat[Correct script to move the sprite five steps whenever the space key is pressed.]{
    \centering
    \includegraphics[width=0.45\columnwidth]{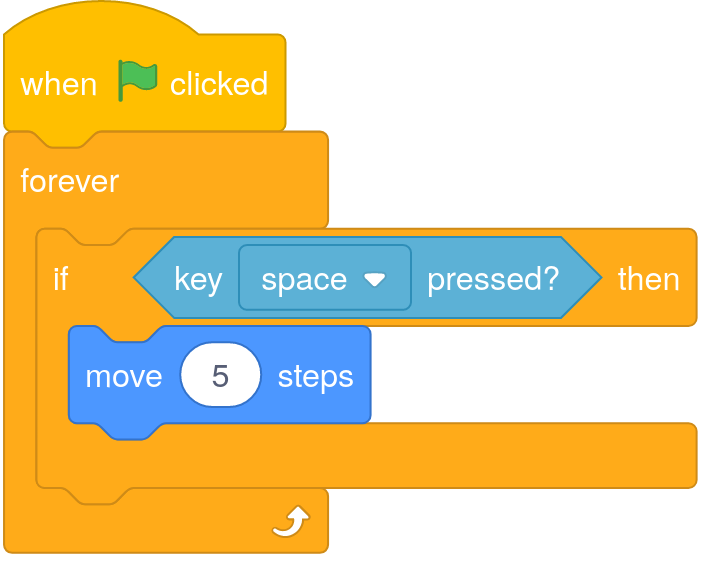}
    \label{fig:loop-sensing}
    }
    \hfill
    \subfloat[Wrong block use: The sprite will
    go to the same position on the stage every time the space key is pressed.]{
    \includegraphics[width=0.45\columnwidth]{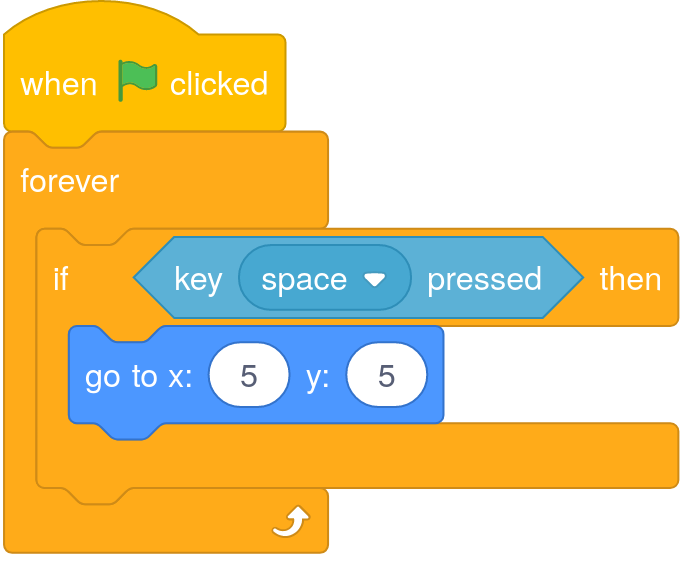}
    \label{fig:go-to-pos}
    }
    \caption{\label{fig:intro}Two scripts aiming to implement the same functionality.}
\end{figure}

Fig.~\ref{fig:loop-sensing} shows a common
programming example in \scratch: the script continuously checks if 
the user has pressed the space key, and whenever this happens 
the sprite 
 is moved 
 by five steps.
Fig.~\ref{fig:go-to-pos} shows a script that tries to accomplish the
same but uses a wrong block: Instead of the \scratchblock{move steps} block,
the \scratchblock{go to position} block is used. Generic linters would miss
this bug as it is project-specific and does not violate any general programming
concepts. Even an automated test only pressing the space key once would
incorrectly report this behavior as correct.
Given a dataset of students' solutions for this task,
anomaly detection learns common patterns such as to use a \scratchblock{move steps} block
whenever the \scratchblock{when green flag}, \scratchblock{forever} and \scratchblock{if-then}
blocks are combined. Consequently, the buggy script in Fig.~\ref{fig:go-to-pos} would be
flagged as an anomaly.\\
In this paper, we introduce the concept of anomaly detection in the classroom.
In detail, the contributions of this paper are:
\begin{itemize}
    \item We formally introduce and implement anomaly detection for \scratch (Section~\ref{sec:anomaly-for-scratch}).
    \item We empirically evaluate the practical applicability of anomaly detection for \scratch (Section~\ref{sec:eval}).
\end{itemize}

%
Evaluated on a dataset of six \scratch
programming projects with many different student solutions,
our implemention of anomaly detection for \scratch
demonstrates that anomaly detection is a reliable way to find
generic defects as well as project-specific ones, such as the one in
Fig.~\ref{fig:intro}, without any manual labor required from
teachers.

%

\section{Background}
\label{ch:classroom}

Since programming knowledge, skills and mental models cannot be
effectively acquired in the abstract, programming education is heavily
based on practical exercises~\cite{Hassinen.2006}. Students typically
implement similar tasks based on textual specifications of what the
programs should achieve, practicing concepts they first learned about
in theory. In the sense of formative and summative assessment, the
results of such tasks can provide educators with clues that they can
use to evaluate and improve the students' learning~\cite{Grover.2020}:
Teachers, tutors, and automated tutoring systems need to interact with
students during assignments to provide feedback and help during
exercise sessions, or to evaluate and grade submissions; this applies
equally to textual and visual programming languages. In this section
we explore what means for support exist in this setting, and how
anomaly detection can help, focusing particularly on the visual
programming language \scratch.

\subsection{Evaluating Student Programs}

In order to teach programming, educators need to have content knowledge (CK) as well as pedagogical content knowledge (PCK)~\cite{Hubwieser.2013}. The latter is required for planning and conducting programming lessons and comprises various aspects that influence the learning process. According to the model of Magnusson et al.~\cite{Magnusson.1999}, PCK includes, amongst others, knowledge about suitable assessment strategies to evaluate students’ understanding. This aspect is particularly important, as studies show that teachers' insufficient understanding of their students reduces the quality of their teaching~\cite{Rahimi.2017}.

While Grover argues for a range of different assessment types during learning to program~\cite{Grover.2017}, the most obvious and common method of assessing programming skills is to evaluate the students' programs~\cite{Insa.2016}. This can provide important insights to educators---e.g. by exposing misconceptions or gaps in the students' understanding---but is particularly challenging for novice or inexperienced teachers~\cite{Sentance.2017}. 

A primary means to support the analysis of learners' code is by running automated tests
against the solutions. The most common application for this is
automated grading: By implementing individual tests for the various
requirements that a program should satisfy, the resulting grade can be
determined as the ratio of tests that a submission passes. 
This general principle 
is implemented in numerous grading tools, which are summarized in various
surveys~\cite{ala2005survey,douce2005automatic,ihantola2010review}.
%
Automated tests can also serve as
feedback to students, or as the basis for producing hints and
corrections~\cite{perelman2014test,gulwani2016automated}.


In practice, a primary challenge for the application of automated
tests lies in their creation. First, it requires the existence of
appropriate automation frameworks in which to specify and execute
these tests, which are not always available. Second,
creating suitable tests is challenging, even for professional
developers~\cite{pham2014enablers,blondeau2017testing,beller2017developer,pecorelli2020testing}.

Static analysis tools are sometimes applied for checking style, code
smells, and bugs in student code. For example, the industrial strength
\toolname{FindBugs}~\cite{hovemeyer2004finding} tool has been
investigated in an educational domain~\cite{edwards2019can}, and can
be integrated into the build tool chain of modern
autograders~\cite{krusche2018artemis}. Such static analysis tools
require no specification effort from the teacher, but the scope of the
feedback they can produce is limited: They can only report generic,
assignment-independent issues.

\subsection{The \scratch Programming Language}

\scratch~\cite{maloney2010} is a block-based programming language.
Programmers can choose from over one hundred
\textit{blocks}\footnote{\url{https://en.scratch-wiki.info/wiki/Blocks},
last accessed February 12, 2021}
which resemble puzzle pieces. 
The blocks can be composed visually with each other 
in the \scratch editor\footnote{\url{https://scratch.mit.edu/projects/editor/}, 
last accessed February 12, 2021}
to define the behavior of \scratch programs.
A collection of blocks that are connected to one unit 
is called a \textit{script}. 
Usually, a script begins with a \textit{hat block} which is an 
event listener. 
The hat block is followed by an arbitrary number of blocks that define 
the actions to execute after the event of the hat block was triggered. 
Scripts belong to \textit{actors}~\cite{VerifiedFromScratch}, 
that is, either the stage or one of the sprites. 
The \emph{stage} is the background of the program; 
\emph{sprites} are the objects acting on the stage.
Fig.~\ref{fig:intro} illustrates two \scratch scripts, both
are triggered by the green-flag event---that is, start executing
when the program starts---by clicking the green flag symbol in the \scratch editor.
More details on \scratch and formalizations thereof can be 
found in the literature~\cite{maloney2010,whisker,VerifiedFromScratch}.

Blocks have different shapes and colors to distinguish between
different categories of statements and expressions, for example, 
event listeners, or control structures. 
Generally, we distinguish between command blocks and reporter blocks.
When executed, a \emph{command block} performs different actions
under specified conditions.
Hat blocks, control blocks, stack blocks, and cap blocks are
types of command blocks.
A \emph{reporter block} describes an expression to evaluate and
produces a scalar value, for example, an integer, Boolean, or a string.


%

\subsection{Program Analysis for \scratch}\label{subsec:background-scratch}

The increasing popularity of \scratch as an introductory programming
environment has triggered research on analyzing the resulting
programs.  In particular, the observation that \scratch programmers
tend to develop certain negative habits while
coding~\cite{meerbaum2011habits} has led to investigations into the
general quality problems in \scratch programs using static analysis
tools.  It has been shown that various types of code smells are
prevalent~\cite{aivaloglou2016kids,hermans2016a,techapaloku2017b, robles2017software}
and have a negative impact on code
understanding~\cite{hermans2016code}.
%
There are tools for finding code smells in \scratch programs such as
\textsc{Hairball}~\cite{boe2013hairball}, \textsc{Quality
  hound}~\cite{techapaloku2017b} or
\textsc{SAT}~\cite{chang2018scratch},
and \litterbox~\cite{bugpatterns} detects predefined bug patterns
automatically.

Testing frameworks have also been proposed for \scratch. In
particular, \textsc{Itch}~\cite{johnson2016itch} translates a small
subset of \scratch programs (say/ask blocks) to Python programs and
then runs tests on these programs.  The \whisker tool~\cite{whisker}
executes automated tests directly in the \scratch IDE, and supports
property-based testing.  \bastet~\cite{VerifiedFromScratch} provides a
general program analysis framework that can be used for any
configurable program analysis, such as software model
checking.

\subsection{Anomaly Detection}

An alternative to the common types of program analysis described above
is offered by the concept of anomaly detection. The general principle is
that likely rules about software projects, programming practices, or
API usage are inferred automatically from source code, version
histories, or execution traces. Violations of these rules (anomalies),
are then reported as likely bugs. The quality of the reported
violations depends on how rules are encoded, the algorithms used for
mining the rules and determining outliers, as well as the
data source. 

There is a variety of technical approaches: Techniques based on
frequent itemset mining techniques capture co-occurrences of methods
and variables~\cite{dynamine,li2005pr}.
%
%
These techniques can be extended to capture control flow information
using graph
models~\cite{eisenbarth,grouminer,wasylkowski2011mining,jadet}. For
example, the \jadet tool~\cite{jadet} extracts temporal properties
that capture common sequences of method calls on instances of \java
classes. An alternative lies in the use of n-gram language models to
capture the regularities of software source code, and then to report
aspects of code with low probabilities as
suspicious~\cite{wang2016bugram}.

A common assumption of these approaches is that anomaly detection is
applied on large software projects, or on large collections of
software projects that share some properties (e.g., common
dependencies), such that the data mining algorithms succeed in
extracting relevant patterns. In contrast, programs in an educational
context tend to be small and on their own do not provide sufficient
opportunity for mining properties.  However, in contrast to a regular
software engineering scenario there is redundancy in terms of multiple
student solutions for the same problem, which we aim to exploit in
this paper.


\section{Anomaly Detection for Scratch}\label{sec:anomaly-for-scratch}

In this section, we describe how anomaly detection can be 
implemented for \scratch programs.
We build on an existing approach that was presented for object-oriented
programs~\cite{jadet, diss} and adjust it for \scratch programs.

\subsection{Modeling Control Flow with Script Models}
We aim to find violations of temporal activations of blocks, 
and therefore model the control flow of \scratch programs. 
The control flow between blocks in a script is 
represented by its \sm, which describes how the 
control of the program execution flow is passed between 
the blocks of a \scratch program.
Formally, we define a \emph{\sm} as follows:

\newcommand{\op}{\ensuremath{b}\xspace}%
\newcommand{\ops}{\ensuremath{B}\xspace}%
\newcommand{\block}{\op\xspace}%
\newcommand{\blocks}{\ops\xspace}%
\newcommand{\script}{\ensuremath{m}\xspace}%
\newcommand{\scriptlist}{\ensuremath{\overline{m}}\xspace}%
\newcommand{\scripts}{\ensuremath{M}\xspace}%
\newcommand{\locations}{\ensuremath{L}\xspace}%
\newcommand{\location}{\ensuremath{l}\xspace}%
\newcommand{\entrylocation}{\ensuremath{\location_0}\xspace}%
\newcommand{\exitlocations}{\ensuremath{\locations_x}\xspace}%
\newcommand{\ctrltransitions}{\ensuremath{G}\xspace}%
\newcommand{\ctrltransition}{\ensuremath{g}\xspace}%
\begin{definition}[Script Model\label{def:sm}]
A \emph{\sm} is a tuple~$\script = (\locations, \ops, \ctrltransitions, \entrylocation, \exitlocations)$, 
with a finite set~\locations of \emph{control locations}, 
a finite set~\ops of \emph{command blocks}, 
a \emph{control transition relation}~$\ctrltransitions \subseteq \locations \times (\ops \cup \{ \epsilon \}) \times \locations$,
an \emph{initial} control location~$\entrylocation \in \locations$,
and a set of control \emph{exit locations}~$\exitlocations \subseteq \locations$.
A control location can be \emph{reached} by executing the blocks on the
transitions in the control transition relation, 
starting from the initial location~\entrylocation.
\end{definition}

\noindent
Epsilon~($\epsilon$) moves are used (1)~for abstracting away
command blocks that are irrelevant for anomaly detection,
and (2)~as a convenience feature to create the script models.
Epsilon elimination as known from $\epsilon$-NFAs~\cite{AutomataTheory}
is applicable.
All definitions that follow assume that script models are  \emph{$\epsilon$-free},
that is, that all $\epsilon$-moves have been eliminated upfront.

\begin{figure}[tp]
\centering
\begin{tikzpicture}[node distance=25mm, scale=0.8, transform shape,
cell/.style={rectangle,draw=black}, initial text={},
space/.style={minimum height=1.2em,matrix of nodes,row sep=-\pgflinewidth,column sep=-\pgflinewidth}]
\node[cfastate,initial] (l0) {$l_0$};
\node[cfastate,right=of l0] (l1) {$l_1$};
\node[cfastate,right=of l1] (l2) {$l_2$};
\node[cfastate,right=of l2] (l3) {$l_3$};
\path[trans] (l0) edge node [pos=.5,label=above:{$\scratchblock{if-then}$}] {} (l1);
\path[trans] (l1) edge node [pos=.5,label=above:{$\scratchblock{key pressed}$}] {} (l2);
\path[trans] (l2) edge node [pos=.5,label=above:{$\scratchblock{move steps}$}] {} (l3);
\end{tikzpicture}
\begin{tikzpicture}[node distance=25mm, scale=0.8, transform shape,
cell/.style={rectangle,draw=black}, initial text={},
space/.style={minimum height=1.2em,matrix of nodes,row sep=-\pgflinewidth,column sep=-\pgflinewidth}]
\node[cfastate,initial] (l0) {$l_0$};
\node[cfastate,right=of l0] (l1) {$l_1$};
\node[cfastate,right=of l1] (l2) {$l_2$};
\node[cfastate,right=of l2] (l3) {$l_3$};
\path[trans] (l0) edge node [pos=.5,label=above:{$\scratchblock{if-then}$}] {} (l1);
\path[trans] (l1) edge node [pos=.5,label=above:{$\epsilon$}] {} (l2);
\path[trans] (l2) edge node [pos=.5,label=above:{$\scratchblock{move steps}$}] {} (l3);
\end{tikzpicture}
\begin{tikzpicture}[node distance=25mm, scale=0.8, transform shape,
cell/.style={rectangle,draw=black}, initial text={},
space/.style={minimum height=1.2em,matrix of nodes,row sep=-\pgflinewidth,column sep=-\pgflinewidth}]
\node[cfastate,initial] (l0) {$l_0$};
\node[cfastate,right=of l0] (l1) {$l_1$};
\node[cfastate,right=of l2] (l3) {$l_3$};
\path[trans] (l0) edge node [pos=.5,label=above:{$\scratchblock{if-then}$}] {} (l1);
\path[trans] (l1) edge node [pos=.5,label=above:{$\scratchblock{move steps}$}] {} (l3);
\end{tikzpicture}
\caption{Abstraction of script models}
\label{fig:scriptabstraction}
\vspace{-0.5em}
\end{figure}
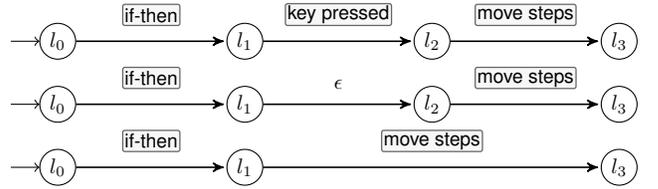

\begin{example}
Fig.~\ref{fig:scriptabstraction} illustrates a script model
and (1)~how particular blocks can be abstracted away by (2)~replacing 
them by $\epsilon$-moves and (3)~eliminating the $\epsilon$-moves in the end.
In this example, the reporter block~\scratchblock{key pressed}
is removed to discover more generic patterns.
\end{example}

\noindent
Note that we generally abstract away reporter blocks in this
work to discover more generic patterns. 

In contrast to a control flow graph or automaton, a script model 
contains transitions that are labeled with control blocks despite 
the fact that the semantics of these blocks is encoded into the graph structure.
Fig.~\ref{fig:intro} provides an example where the control
block \scratchblock{forever} must precede the control block \scratchblock{if-then}.

\begin{figure}[t]
   \centering
   \subfloat[Correct:
    Forever loop resulting in an infinite sequence of moves.]{
       \centering
\begin{tikzpicture}[node distance=6mm, scale=0.8, transform shape,
cell/.style={rectangle,draw=black}, initial text={},
space/.style={minimum height=1.2em,matrix of nodes,
row sep=-\pgflinewidth,column sep=-\pgflinewidth}]
\node[cfastate,initial] (l0) {$l_0$};
\node[cfastate,below=of l0] (l1) {$l_1$};
\node[cfastate,below=of l1] (l2) {$l_2$};
\node[cfastate,below right=of l2] (l3) {$l_3$};
\path[trans] (l0) edge node [pos=.5,label=right:{$\scratchblock{when green flag}$}] {} (l1);
\path[trans] (l1) edge node [pos=.5,label=right:{$\scratchblock{forever}$}] {} (l2);
\path[trans] (l2) edge [loop left] node [pos=.1, label=left:{$\scratchblock{if-then}$}] {} (l2);
\path[trans] (l2) edge [bend left] node [pos=.5, label=right:{$\scratchblock{if-then}$}] {} (l3);
\path[trans] (l3) edge [bend left] node [pos=0, label=left:{$\scratchblock{move
steps}$}] {} (l2);
\end{tikzpicture}
       \label{fig:loop-aum}
   }
   \hspace*{0.1cm}
   \subfloat[Buggy: Differs in the block
   used between~$l_3$ and~$l_2$.]{
       \centering
\begin{tikzpicture}[node distance=6mm, scale=0.8, transform shape,
cell/.style={rectangle,draw=black}, initial text={},
space/.style={minimum height=1.2em,matrix of nodes,
row sep=-\pgflinewidth,column sep=-\pgflinewidth}]
    \node[cfastate,initial] (l0) {$l_0$};
    \node[cfastate,below=of l0] (l1) {$l_1$};
    \node[cfastate,below=of l1] (l2) {$l_2$};
    \node[cfastate,below right=of l2] (l3) {$l_3$};
    \path[trans] (l0) edge node [pos=.5,label=right:{$\scratchblock{when green flag}$}] {} (l1);
    \path[trans] (l1) edge node [pos=.5,label=right:{$\scratchblock{forever}$}] {} (l2);
    \path[trans] (l2) edge [loop left] node [pos=.1, label=left:{$\scratchblock{if-then}$}] {} (l2);
    \path[trans] (l2) edge [bend left] node [pos=.5, label=right:{$\scratchblock{if-then}$}] {} (l3);
    \path[trans] (l3) edge [bend left] node [pos=0, label=left:{$\scratchblock{go to position}$}] {} (l2);
\end{tikzpicture}
       \label{fig:missing-loop-aum}
   }
   \caption{\label{fig:aum-intro} The script models of the scripts in Fig.~\ref{fig:intro}. 
    Nodes are locations in the code, outgoing transitions are labeled with the 
    blocks that can be executed from this location. 
    The script models show command blocks and abstract away block
   inputs.}
\vspace{-0.5em}
\end{figure}
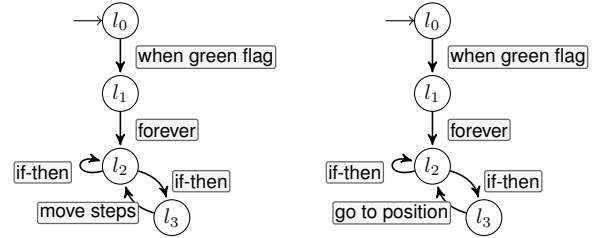

\begin{example}
Fig.~\ref{fig:aum-intro} shows the \sms of the \scratch scripts from
Fig.~\ref{fig:intro}.
The nodes of this graph refer to control locations in a script, 
and the edges in between them denote blocks that can be executed 
from these locations. 
\end{example}

\subsection{Extracting Block Patterns from Script Models}
\newcommand{\propsof}{\ensuremath{\mathsf{props}}\xspace}%
\newcommand{\andlater}{\propsof}%


Every single script~(represented by a script model) implements a set 
of temporal properties that define how the script behaves over time.
In a later step, behavioral patterns are mined by analyzing the 
temporal properties of a large set of script models---%
in contrast to related work~\cite{jadet}, we do not use object usage models.
Before we define the notion of temporal properties, we define
the transitive closure of a script model:

\begin{definition}[Transitive Closure]
Given an $\epsilon$-free script model~$\script = (\locations, \ops, \ctrltransitions, \entrylocation, \exitlocations)$, 
we define 
the transitive closure~$\ctrltransitions^+ \subseteq \locations \times \blocks \times \locations$
of its control transition relation~\ctrltransitions
recursively as~$\ctrltransitions^+ = \ctrltransitions \cup \{ (\location_1, \block, \location_3) \; 
| \; (\location_1, \block, \location_2), (\location_2, \block, \location_3) \in \ctrltransitions^+ \}$.
\end{definition}

\begin{definition}[Temporal Properties~\cite{jadet}]
The \emph{temporal property relation}~$\prec \; \subseteq \blocks \times \blocks$
of a script~$\script \in \scripts$ defines the pairs of blocks 
that occur one after the other in its control flow,
possibly interleaved with the execution of other blocks.
That is, $\prec \; = \{ (\block_1, \block_2) \; | \;
(\cdot, \block_1, \location_1) \in \ctrltransitions^+ 
\land (\location_1, \block_2, \cdot) \in \ctrltransitions^+ \}$.
We write~$\block_1 \prec \block_2$ if and only if
$(\block_1, \block_2) \in \prec$.
We use the alternative notation~$\propsof(\script) \subseteq \blocks \times \blocks$ 
to denote the temporal properties of a given script~\script.
\end{definition}

\noindent
In other words, the temporal property relation is defined by the blocks 
that we can reach eventually in the \sm starting from a block at hand. 

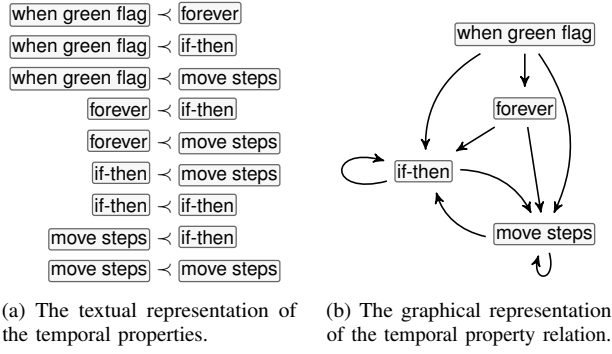
\begin{figure}
    \centering
    \subfloat[The textual representation of the temporal properties.]{%
\begin{tikzpicture}[node distance=6mm, scale=0.8, transform shape,
cell/.style={rectangle,draw=black}, initial text={},
space/.style={minimum height=1.2em,matrix of nodes,
row sep=-\pgflinewidth,column sep=-\pgflinewidth}]
\node[] (l0) {
    $\begin{aligned}%
        \scratchblock{when green flag} &\prec \scratchblock{forever} \\
        \scratchblock{when green flag} &\prec \scratchblock{if-then} \\
        \scratchblock{when green flag} &\prec \scratchblock{move steps} \\
        \scratchblock{forever} &\prec \scratchblock{if-then} \\
        \scratchblock{forever} &\prec \scratchblock{move steps} \\
        \scratchblock{if-then} &\prec \scratchblock{move steps} \\
        \scratchblock{if-then} &\prec \scratchblock{if-then} \\
        \scratchblock{move steps} &\prec \scratchblock{if-then} \\
        \scratchblock{move steps} &\prec \scratchblock{move steps} \\
    \end{aligned}$};
\end{tikzpicture}
    }%
\hspace{4mm}%
    \subfloat[The graphical representation of the temporal property relation.]{%
        \centering
\begin{tikzpicture}[node distance=6mm, scale=0.8, transform shape,
cell/.style={rectangle,draw=black}, initial text={},
space/.style={minimum height=1.2em,matrix of nodes,
row sep=-\pgflinewidth,column sep=-\pgflinewidth}]
\node[] (l0) {\scratchblock{when green flag}};
\node[below=of l0] (l1) {\scratchblock{forever}};
\node[below left=of l1] (l2) {\scratchblock{if-then}};
\node[below right=of l2] (l3) {\scratchblock{move steps}};
\path[trans] (l0) edge node [pos=.5,label=right:{}] {} (l1);
\path[trans] (l0) edge [bend right] node [pos=.5,label=right:{}] {} (l2);
\path[trans] (l0) edge [bend left] node [pos=.5,label=right:{}] {} (l3);
\path[trans] (l1) edge node [pos=.5,label=right:{}] {} (l2);
\path[trans] (l1) edge node [pos=.5,label=right:{}] {} (l3);
\path[trans] (l2) edge [bend left] node [pos=.5,label=right:{}] {} (l3);
\path[trans] (l3) edge [bend left] node [pos=.5,label=right:{}] {} (l2);
\path[trans] (l2) edge [loop left] node [pos=.5,label=right:{}] {} (l2);
\path[trans] (l3) edge [loop below] node [pos=.5,label=right:{}] {} (l3);
\end{tikzpicture}
        \label{fig:pattern-img}
    }
    \caption{\label{fig:properties}The temporal properties of 
    the \sm in Fig.~\ref{fig:loop-aum}.}
\end{figure}

\begin{example}
Using the temporal property relation
we can now analyze pairs~$(\block_1, \block_2) \in \prec$ of blocks,
where one block~$\block_1$ may precede the other block~$\block_2$. 
Fig.~\ref{fig:properties} shows the temporal property
relation for the \sm illustrated in Fig.~\ref{fig:loop-aum}.
\end{example}

We use the notion of patterns to learn about common
temporal behavior of scripts (and their models),
which is central for detecting anomalies (deviations
from common patterns).

\newcommand{\pattern}{\ensuremath{p}\xspace}%
\newcommand{\blockpatterns}{\ensuremath{P}\xspace}%
\begin{definition}[Pattern~\cite{jadet}]
A \emph{pattern}~$\pattern \subseteq \ops \times \ops$ is a set of 
temporal properties, where one temporal property is a pair of blocks.
A pattern~$\pattern$ is \emph{supported} by a script~$\script$ if 
$\pattern$ defines a subset of its temporal properties,
that is, if~$\pattern \subseteq \; \propsof(\script)$.
The set of all possible patterns is denoted by the symbol~$\blockpatterns$.
\end{definition}

\newcommand{\supportof}{\ensuremath{\mathsf{supp}}\xspace}%
\newcommand{\support}{\ensuremath{s}\xspace}%
\begin{definition}[Pattern Support~\cite{jadet}]
Given a list of scripts~$\scriptlist = \langle \script_1, \ldots, \script_n \rangle$,
the \emph{support}~$\supportof(\pattern, \scriptlist) \rightarrow \mathbb{N}_0$ 
of a pattern~\pattern is the number of scripts that support the pattern, that is, 
$\supportof(\pattern, \scriptlist) = | \{ \script \; | \; \pattern \subseteq \propsof(\script) \land \; \script \in \scriptlist \} |$.
\end{definition}

\begin{figure}
    \centering
\begin{tikzpicture}[node distance=6mm, scale=0.8, transform shape,
cell/.style={rectangle,draw=black}, initial text={},
space/.style={minimum height=1.2em,matrix of nodes,
row sep=-\pgflinewidth,column sep=-\pgflinewidth}]
\node[] (l0) {\scratchblock{when green flag}};
\node[below=of l0] (l1) {\scratchblock{forever}};
\node[below left=of l1] (l2) {\scratchblock{if-then}};
\node[right=of l2] (l3) {\mscratchblock{move steps}};
\path[trans] (l0) edge node [pos=.5,label=right:{}] {} (l1);
\path[trans] (l0) edge [bend right] node [pos=.5,label=right:{}] {} (l2);
\path[missing] (l0) edge [bend left] node [pos=.5,label=right:{}] {} (l3);
\path[trans] (l1) edge node [pos=.5,label=right:{}] {} (l2);
\path[missing] (l1) edge node [pos=.5,label=right:{}] {} (l3);
\path[missing] (l2) edge [] node [pos=.5,label=right:{}] {} (l3);
\path[missing] (l3) edge [bend left] node [pos=.5,label=right:{}] {} (l2);
\path[trans] (l2) edge [loop left] node [pos=.5,label=right:{}] {} (l2);
\path[missing] (l3) edge [loop right] node [pos=.5,label=right:{}] {} (l3);
\end{tikzpicture}
    \caption{\label{fig:violation-loop-sensing} Comparison between the temporal properties of
    the scripts in Figs.~\ref{fig:loop-sensing} and~\ref{fig:go-to-pos}.
    The temporal property relation of the buggy script does not contain the 
    properties related to the missing move block and therefore violates
    the pattern of the correct script. Missing properties
    are depicted red and dotted.}
\vspace{-0.5em}
\end{figure}
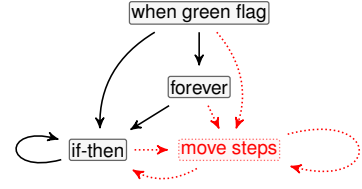

\begin{example}
Consider the list~$\scriptlist = \langle \script_1, \script_2 \rangle$ of script models,
which correspond to the scripts illustrated in Fig.~\ref{fig:intro}.
When considering the set of the temporal properties in Fig.~\ref{fig:properties}
as one pattern, this pattern has support~\num{1} based on the scripts~$\scriptlist$.
The script in Fig.~\ref{fig:loop-sensing} adheres to every 
temporal property of this pattern, whereas the script in Fig.~\ref{fig:go-to-pos}
does not exhibit several of the temporal properties of the pattern.
Fig.~\ref{fig:violation-loop-sensing} shows the missing temporal properties
of the script---indicated with the color red and dotted lines.
As the script does not have a \scratchblock{move steps} block, all the temporal properties
containing the block \scratchblock{move steps} are missing.
Therefore, the buggy script does not support the pattern and it has a support of 1.
\end{example}

Even though script models and block patterns are closely related and their
graphical representation is similar, there are some key differences:
The level of abstraction of patterns is higher than the level 
of abstraction of script models. While a script model only abstracts away reporter blocks,
and therefore represents a limited set of scripts, there is an unlimited 
variety of scripts that may support a pattern.
For example, a temporal property like ``\scratchblock{if-then} $\prec$ \scratchblock{if-then}''
is supported by both a script in which a single \scratchblock{if-then} block 
occurs in a loop and a script in which there are two directly consecutive
\scratchblock{if-then} blocks.

The set of actual patterns found in a set of script models (with corresponding
temporal property relations) is computed using frequent itemset mining:

\newcommand{\freqitemsets}{\ensuremath{\mathsf{freq}}\xspace}%
\newcommand{\nats}{\ensuremath{\mathbb{N}}\xspace}%
\newcommand{\minsupport}{\ensuremath{k}\xspace}%
\begin{definition}[Frequent Itemsets~\cite{jadet}]
Frequent itemset mining~$\freqitemsets: 2^{2^{\blocks \times \blocks}} \times \nats \rightarrow 2^{2^{\blocks \times \blocks}}$
takes a set of sets of temporal properties and a minimum 
support threshold~$\minsupport \in \nats$ as argument and 
produces a set of patterns 
that occur
in at least \minsupport sets.
\end{definition}


\subsection{Violations of Block Patterns}\label{sec:violations-block-patterns}

Based on the concepts that we have described in previous sections,
we now discuss how we identify anomalies in \scratch programs.
Anomaly detection can help to show the absence of functionality.
Note that anomaly detection is performed on closed patterns only,
which are defined as follows:

\begin{definition}[Closed Pattern~\cite{jadet}]
A pattern is called \emph{closed} if each pattern that 
is a superset has less support.
\end{definition} 

\newcommand{\violationsof}{\ensuremath{\mathsf{viol}}\xspace}%
\begin{definition}[Violation~\cite{jadet}]
A script~$\script$ \emph{violates} a pattern~$\pattern$ 
if the pattern is not a subset of the temporal properties of the script,
that is, if and only if $\pattern \not \subseteq \propsof(\script)$.
\end{definition}

\noindent
Violations hint at scripts that do not support every temporal property 
of a common pattern. 
Therefore, the violation of a block pattern always consists of two sets
of temporal properties: A set of sequential constraints which are 
adhered to, and a set of missing temporal properties---the \textit{deviation}.

\newcommand{\deviation}{\ensuremath{\mathsf{devi}}\xspace}%
\begin{definition}[Deviation~\cite{diss}]
Given a script~$\script$ and a pattern~$\pattern$, the \emph{deviation} is the set of
temporal properties  $\deviation(\script, \pattern) = \pattern \setminus \propsof(\script)$
that are missing in the script.
\end{definition}

\begin{example}
Fig.~\ref{fig:violation-loop-sensing}, which compares the temporal properties
of the scripts in Fig.~\ref{fig:intro}, shows the violation of the buggy script.
The \textit{deviation} consists of all five temporal properties related
to the \scratchblock{move steps} block.
\end{example}

Not all violations hint at defects or contribute new knowledge.
The \textit{confidence value of a block pattern violation} is defined 
by the confidence of its deviation and measures how many scripts 
exhibit the exact same deviation from the same pattern the violation violates.

\newcommand{\violations}{\ensuremath{v}\xspace}%
\newcommand{\confidence}{\ensuremath{c}\xspace}%
\begin{definition}[Violation Confidence~\cite{jadet}]
Given a list of scripts~$\scriptlist = \langle \script_1, \ldots, \script_n \rangle$,
a script $\script$ and a pattern~\pattern, the \emph{confidence}
of a violation of pattern~\pattern of script \script is the
ratio~$\confidence = \support / (\support + \violations)$,
with the support~$\support = \supportof(\pattern, \scriptlist)$
and the number of violations that violate $\pattern$ the same way $\script$ does:
$\violations = | \{m_i \; | \; \deviation(m_i, \pattern) = \deviation(\script, \pattern) \land m_i \in \scriptlist\} |$.
\end{definition}

\begin{definition}[Anomaly~\cite{jadet}]
An \emph{anomaly} is a violation of a block pattern by
a script for that the violation confidence is above a 
particular threshold~(minimum confidence).
\end{definition}

\noindent
The actual identification of anomalies is implemented
based on Formal Concept Analysis.
A lattice of closed patterns is traversed from the top
element (the pattern with the highest support) down to
elements with lower support (until a min-support 
limit is reached)~\cite{jadet}. The anomalies found are
ranked and filtered using methods from Association Rule
Mining to report anomalies likely pointing at erroneous
behaviour~\cite{diss}.

\subsection{Implementation}
\label{sec:impl}%
%

Our tool chain for anomaly detection for \scratch uses an extended version of  \litterbox~\cite{bugpatterns}  to generate a 
collection~$\langle \script_1, \ldots, \script_n \rangle$ of script models for a collection of \scratch projects. These script models are handed over to \jadet~\cite{jadet}
to mine patterns and check for violations.
\jadet's algorithms for pattern and violation mining are not \java-specific: 
This allowed us to adapt \jadet to check \scratch code without algorithmic adaptations.
Note that while \jadet was designed to operate on object usage models
to check for correct API usage, we use script models 
that are not restricted to code that interacts with particular variables or objects.

\subsection{Application}

We envision that a primary application for anomaly detection is to support teachers during formative assessment:
A major advantage of anomaly detection is that it highlights noteworthy or problematic behavior without requiring a detailed and laborious inspection of all student programs. It therefore seems particularly suitable also for real-time feedback during programming classes.
Anomaly detection could similarly support summative assessment, although teachers would in this case need to be particularly aware that \emph{common} erroneous behavior does not represent anomalies. Besides a general understanding of what an anomaly is, however, no further training should be required in order to use anomaly detection in the classroom.
It is also conceivable that anomaly detection could be integrated into hint generation techniques, such that students receive feedback automatically, without the need for teacher interaction. In this context, richer data, for example using a history of previous solutions to the task at hand, could help to improve the quality of reported anomalies.


\section{Empirical Evaluation}\label{sec:eval}

\newcommand{\toollink}{\small{\texttt{\url{https://github.com/se2p/scratch-anomalies}}}}%

To investigate the practical applicability of anomaly
detection in \scratch, we
 aim to empirically answer the following
research questions:
\smallskip
\begin{compactdesc}
    \item \textbf{RQ1} Can anomalies be found in assignment solutions?
    \item \textbf{RQ2} Do erroneous solutions lead to more anomalies?
    \item \textbf{RQ3} Which categories of anomalies can be identified?
\end{compactdesc}

\smallskip
We implemented our approach as an extension of \litterbox~\cite{bugpatterns} and
\jadet~\cite{jadet, diss} and it is available at:

\begin{center}
\toollink
\end{center}

\subsection{Datasets}\label{subsec:datasets}

\begin{table}[t]
  \centering
  \caption{\label{tab:projects}Statistics describing the evaluation datasets}
\begin{tabular}{l@{~}r@{~}r@{~}r@{~}r@{~}r@{~}r}
  \toprule
  Project  & Solutions & $\varnothing$ Blocks & $\varnothing$ Statements & $\varnothing$ Scripts & $\varnothing$ Sprites & WMC \\ 
  \midrule
  Monkey   & 130 & 5.48 & 4.56 & 2.03 & 2.06 & 2.83 \\ 
  Elephant & 130 & 9.45 & 9.40 & 1.18 & 1.16 & 1.87 \\ 
  Cat      & 129 & 7.57 & 5.82 & 3.08 & 1.99 & 5.51 \\ 
  Horse    &  73 & 3.70 & 2.89 & 1.10 & 1.10 & 1.86 \\ 
  Fruit    &  42 & 54.26 & 38.50 & 6.86 & 3.05 & 16.57 \\ 
  Open     & 295 & 34.45 & 28.61 & 7.38 & 4.37 & 13.92 \\ 
   \bottomrule
\end{tabular}
\end{table}

\ignore{R1: Could you provide a link to the data (the student programs they analyzed), as well?}

We use a dataset consisting of student solutions for six different programs:
\begin{itemize}
\item \monkey: The aim of this program is to make the sprite of
  a circus director continuously move towards a monkey~\cite{katharina}.
\item \elephant: The aim of this program is to simulate a
  dancing elephant by continuously switching its costumes (i.e.,
  images representing different poses)~\cite{katharina}.
\item \cat: A cat sprite should indicate with a speech bubble
  whenever it catches the ball~\cite{katharina}.
\item \horse: A horse sprite should continuously change color,
  but when it touches the mouse pointer it should rotate~\cite{katharina}.
\item \fruit: The player controls a fruit bowl
  with the cursor keys, and has to catch fruit dropping down from the top~\cite{whisker}.
\item \open: For this dataset, the students first implemented
  three tightly specified tasks for training, before they were asked
  to implement something similar to the previous tasks, but were not
  given any further specification of what program specifically to
  create. Thus, unlike the other projects, this is an \emph{open} task
  and there is no specification.
\end{itemize}

\done{R1: In their experiment, did the authors characterize the students -- like grade level or age?  Or whether the teaching skills of their teacher could be classified as "typical" for that type of student?}
For each of these tasks we collected student solutions during programming 
sessions conducted by qualified teachers. For the \monkey, \elephant, \cat, and \horse tasks 
solutions were produced by primary school children aged $8$--$10$, 
the \open task was solved by children aged $9$--$12$, and 
the \fruit task was solved by children aged $12$--$13$.
The numbers of solutions as well as size and complexity metrics
are stated in Table~\ref{tab:projects}. Note that we use the
full datasets including empty projects of students
who did not engage at all, since this also represents the actual use
case of a teacher applying our approach.

\subsection{Anomaly Mining}

To mine violations, we extract the script models for each of the six
datasets, and then use \jadet to mine violations.  

\subsubsection*{Extracted Script Models}
Table~\ref{tab:mining-results} shows the number of projects and 
the resulting \sms mined for every task. 
The creation process finished in less than two seconds for every dataset.  
All experiments on our datasets were conducted on an 
off-the-shelf laptop computer as would be available to teachers.

\subsubsection*{Mining Parameters}
\jadet offers four parameters to configure violation mining: The minimum
support and minimum size of a violated pattern, the maximum deviation
level of violations, and the minimum confidence.
For minimum size and maximum deviation level, we fixed the values at
the defaults used by \jadet: The minimum size of a violated pattern
was set to~\num{2} as we are interested in violations independently of
their size, and~\num{10000} for the maximum deviation level as we are
interested in all violations, no matter how many temporal properties
are missing.


\subsection{Experiments}

We conducted several experiments to answer the questions:
\smallskip

\subsubsection*{RQ1} 
To answer RQ1, we computed statistics on the 
script models extracted, as well as patterns and violations reported by \jadet.
Since the chosen approach to anomaly mining has not been used in this
context before, it is unclear what parameter values are best for the
minimum support and minimum confidence. We therefore conducted a
\emph{sensitivity analysis} on these two parameters with minimum size~($=\num{2}$)
and maximum deviation level~($=\num{10000}$) as fixed variables, changing
only minimum support and minimum confidence.  For the minimum support
we tested the values $\langle 1, 5, 10, 15, 20 \rangle$, where $20$ is the default
\jadet value. For confidence we tested the values~$\langle 0.1, 0.2, \ldots, 0.9 \rangle$. 
Intuitively, larger values for both parameters are expected to
produce higher quality anomalies; however, if the values are too large
then there is a risk of missing relevant anomalies. Assuming a
teaching scenario, we thus choose the configuration with the highest
possible values that reports at least~\num{10} anomalies for each dataset.

\begin{figure*}[t]
  \includegraphics[width=\textwidth]{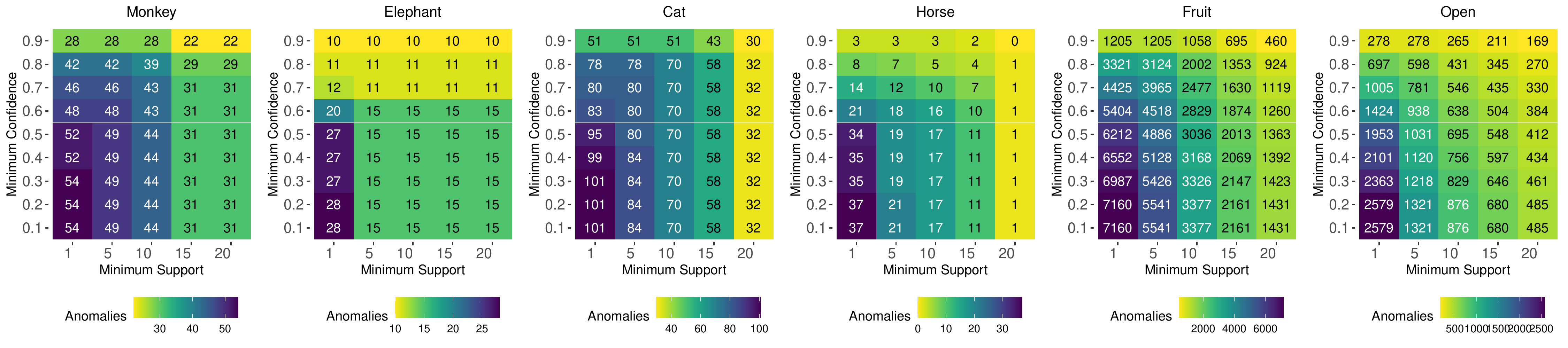}
  \caption{\label{fig:tuning}Tuning results: Numbers of anomalies reported 
    for each of the datasets with different configurations}
\vspace{-0.5em}
\end{figure*}

\subsubsection*{RQ2} 
To answer RQ2, we investigated how correctness of programs relates to 
whether anomalies are reported. We used a
manual classification~\cite{VerifiedFromScratch} of the \monkey,
\cat, \elephant, and \horse datasets, for which the
programs are small enough to allow a binary correct/incorrect
classification; only non-empty projects were classified. For the \fruit dataset, we used the number of
failed tests of the grading test suite used in prior
work~\cite{whisker} as a measurement of the degree of correctness, and
correlated this to the number of anomalies reported. For the
\open dataset a classification in correct/incorrect is not
possible, since there was no specification.

\subsubsection*{RQ3} 
To answer RQ3 we manually classified the top-\num{10} violations reported 
for each of the datasets. Two authors of the paper independently classified 
each of the violations as either:

\done{R1: Were the subcategories (Bug pattern, etc.) chosen prior to inspection of at least some of the examples of code?  I.e., Was there testing of unviewed data (including additional programs) that followed the subcategorizations by the researchers?}

\begin{itemize}
\item \textit{Defective:} 
    The violation hints at a defect in a script that stops it 
    from working in the intended way.
\item \textit{Smelly:} 
    The violation hints at a script that has quality issues but does 
    not break the functionality of the program.
\item \textit{Non-defective:} 
    Adherence to the violated pattern would not contribute to the 
    functionality or quality of the program.
\end{itemize}

To support objective classification, we agreed on subcategories for
every category above, by following the principles of Qualitative 
Content Analysis~\cite{Mayring.2015}:
One author inspected all violations to classify and inductively 
developed subcategories on different levels: Specific subcategories
of the above and more abstract subcategories, 
moving away from script and violation details.
We discussed the resulting abstract subcategories with all the 
authors and agreed on the following subcategories:

\begin{itemize}
    \item \textit{Bug pattern (defective)}: The violation hints at a
     defect that a generic \scratch linter such as
     \litterbox~\cite{bugpatterns} could find equally well. 
    \item \textit{Missing block (defective)}: 
    The violation hints at a missing project-specific block.
    \item \textit{Wrong order of blocks (defective)}: 
    The violation hints at a script with the right blocks 
    assembled in the wrong order.
    \item \textit{Unnecessary block(s) (smelly)}: 
    The violation hints at blocks which are unnecessary, but 
    do not change the functionality of the program.
    \item \textit{Distinguished work (non-defective)}: 
    The violation does not hint at defects or smells in a script.
\end{itemize}

During independent classification by two of the authors, we inspected
the full \scratch program only if the script itself would not provide
sufficient information. 
We classified every violation into one of the subcategories. Examples
for the subcategories are shown in Section~\ref{subsec:rq3}.

\subsection{Threats to Validity}
\label{sec:threats-to-validity}%
As \jadet was left unchanged in all areas that affect the correctness
of the results, the main threat to internal validity is our own
process that extracts the script models. To mitigate this threat, we
wrote automated tests to validate the correctness of the \sms it
creates, and manually inspected a large number of \sms in the
development and classification process.
%
To avoid bias in the manual classification process, we agreed on 
subcategories for every main category, for example,
\textit{bug pattern} is a subcategory of \textit{defective}. 
In the classification process, we assigned both the main
categories and the---less subjective---subcategories to every violation. 
In addition, every violation was classified by two authors and 
divergent assignments were discussed and resolved.
Threats to external validity arise from our choice of parameters as well as the datasets used.
%
We evaluated the effects of the parameters on quantity and quality, but further studies will be necessary to identify parameters that are acceptable for users.
%
Besides the parameters, the quality of violations depends on various properties of the dataset
it is applied to, such as the quality of submissions or sizes, and
%
%
our findings may not generalize to other datasets. However, our
dataset covers different scenarios in terms of class sizes as well as
programming tasks, and we explicitly included closed as well as open tasks.


\subsection{RQ1: Can anomalies be found in solutions?}

Whether or not anomalies can be detected heavily depends on the
parameters of the mining procedure.
To find an appropriate parameterization to analyze script models
extracted from \scratch programs, we conducted a sensitivity
analysis; the results are shown in Fig.~\ref{fig:tuning}.
Based on this analysis, we chose a minimum support of~\num{20} and
a minimum confidence of~\num{0.9} for all datasets except 
the \horse example, where there are fewer solutions and we 
therefore used minimum support~\num{10} and minimum confidence~\num{0.7}.

\begin{figure}[t]
  \centering
  \includegraphics[width=\columnwidth]{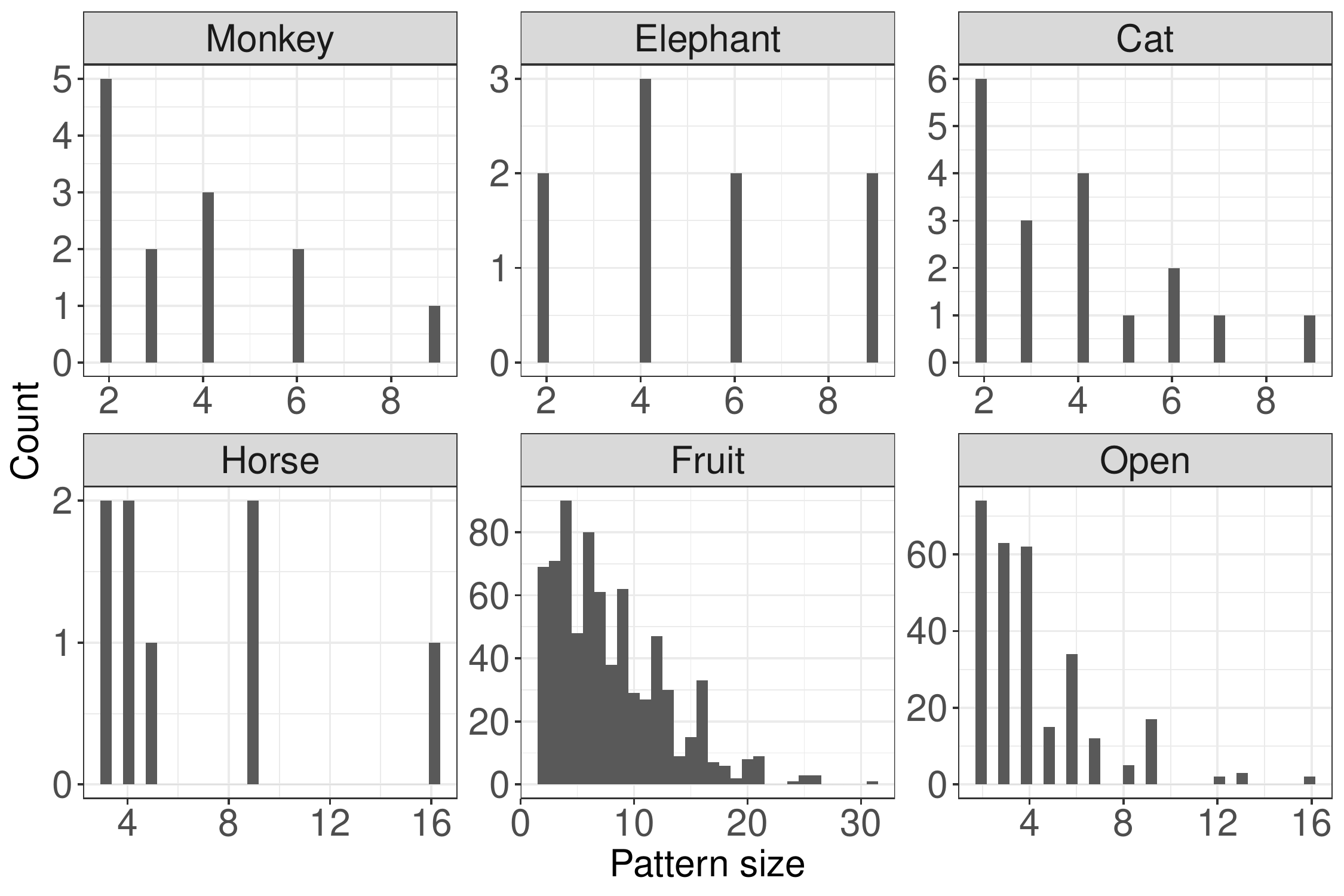}
  \caption{Pattern size distributions of our datasets}
  \label{fig:pattern-size-density}
\vspace{-0.5em}
\end{figure}

\begin{figure}[t]
  \centering
  \includegraphics[width=\columnwidth]{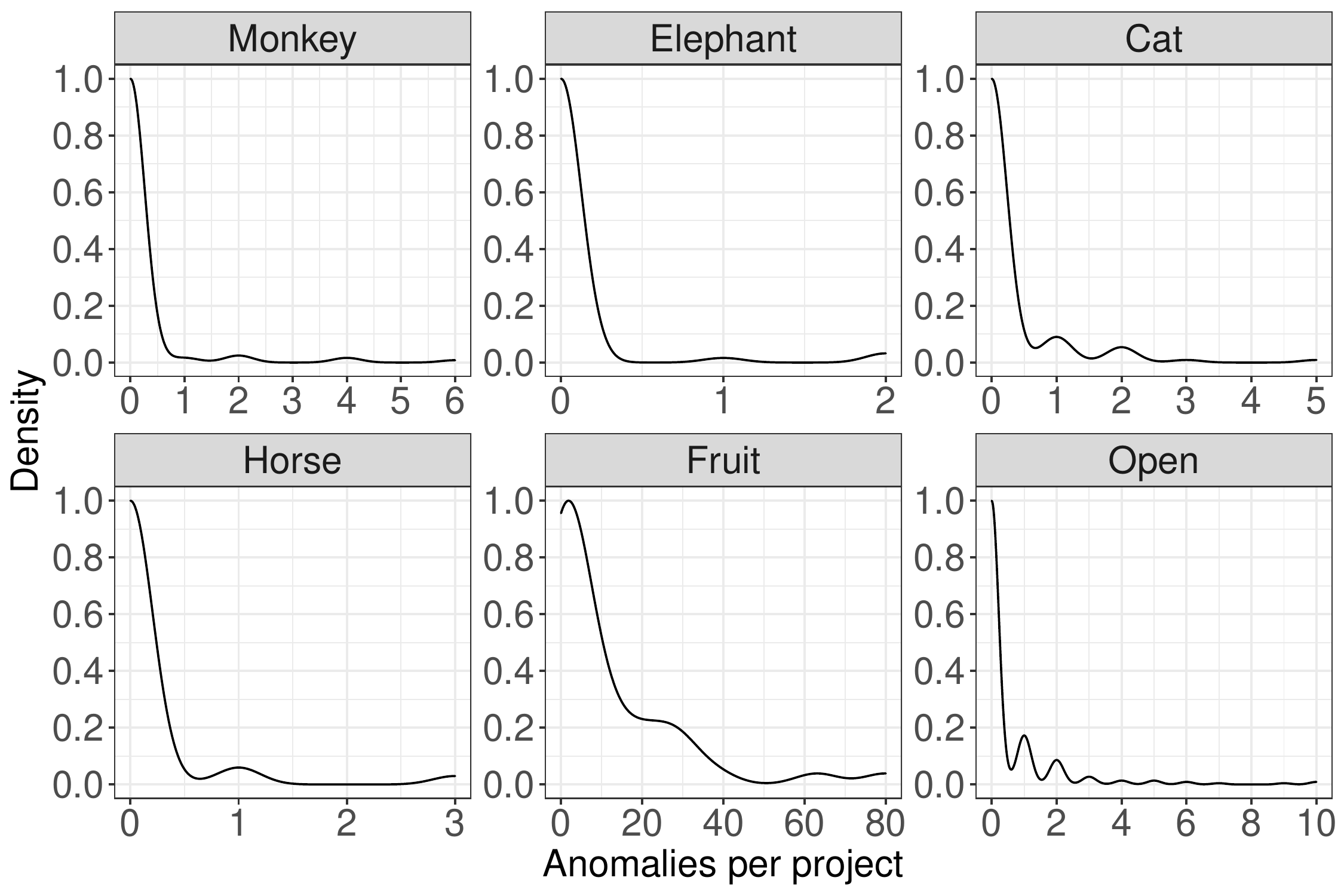}
  \caption{Anomaly distributions of our datasets}
  \label{fig:anomaly-density}
\vspace{-0.5em}
\end{figure}

\begin{table}[t]
    \centering
    \caption{\label{tab:mining-results}Summarized characteristics of the solutions, by task}
    \begin{tabular}{l@{}r@{\phantom{aa}}r@{\phantom{aa}}r@{\phantom{aa}}r@{\phantom{aa}}r}
        \toprule
        Project  & \#Solutions & \#Models & \#Patterns & \#Violations & \#Anomalies\\
        \midrule
        Monkey   & 130 & 264  & 13  & 65   & 22 \\
        Elephant & 130 & 154  & 9   & 34   & 10 \\
        Cat      & 129 & 446  & 18  & 91   & 30 \\
        Horse    & 73  & 80   & 8   & 39   & 10 \\
        Fruit    & 42  & 295  & 749 & 1414 & 460 \\
        Open     & 295 & 2207 & 289 & 684  & 169 \\
        \bottomrule
    \end{tabular}
\end{table}

Table~\ref{tab:mining-results} summarizes the results of the model
extraction and anomaly detection for the chosen parameterization. 
The number of models derived for each of the programs depends on 
the number of scripts in the solutions, and is thus roughly proportional 
to the number of scripts in the solutions as described in Table~\ref{tab:projects}, 
with \cat, \fruit, and \open resulting in the most models. The number of patterns
extracted is lower than the number of models in all but the \fruit
example. For the \open example, the lower number of patterns is
expected since there is more variety in the solutions, as students were
free to implement their own ideas. In the \fruit game, on the other
hand, all students implemented the identical game. In contrast to the
other four closed examples, there is some redundancy within
the scripts in each project, as the behavior of the apple and the
banana sprites share several aspects---both drop from random
locations at the top of the stage to the bottom and check whether they
touch the bowl or the bottom. This shared behavior contributes to the
number of patterns found.

Fig.~\ref{fig:pattern-size-density} summarizes the sizes of these
patterns. The majority of patterns are small, with only few temporal
properties, although all projects have patterns of up to at least nine
properties.  The larger \fruit task stands out with substantially
larger patterns than all other tasks. This is mainly a result of the
overall size and complexity of the projects---see WMC and Statements in
Table~\ref{tab:projects}. Although the \open task contains fairly
complex solutions, too, there is less overlap between these solutions,
resulting in generally smaller patterns.

These patterns tend to lead to multiple violations, as shown in
Table~\ref{tab:mining-results}. However, only a subset of these
violations are reported as mentioned in Section~\ref{sec:violations-block-patterns}. The number of
anomalies generally is roughly proportional to the number of patterns,
ranging from the configured minimum of \num{10} (\elephant) to
\num{460} (\fruit). The number of projects that exhibit
anomalies seems to directly depend on the number of patterns
extracted: For the \elephant, \horse, \monkey, and \cat tasks, less than
\num{15}{\%} of the projects had at least one anomaly. The \fruit task
again stands out with more than half of the projects having reported
at least one anomaly. For \num{25}{\%} of the \open task solutions,
at least one anomaly is reported.  Fig.~\ref{fig:anomaly-density}
summarizes the distribution of anomalies over projects; most projects
have only few anomalies reported, although the \fruit task is the
exception with up to \num{80}~anomalies reported for a single project.

\summary{RQ1}{The number of anomalies reported depends on the 
configuration of the mining process. Our final configuration yields between
\num{10} and \num{460} anomalies per task.
}

\subsection{RQ2: Do erroneous solutions lead to more anomalies?}

\begin{figure}[t]
  \centering
      \subfloat[\label{fig:reported-on-correct}Anomalies reported.]{
      \centering
        \includegraphics[width=0.46\columnwidth]{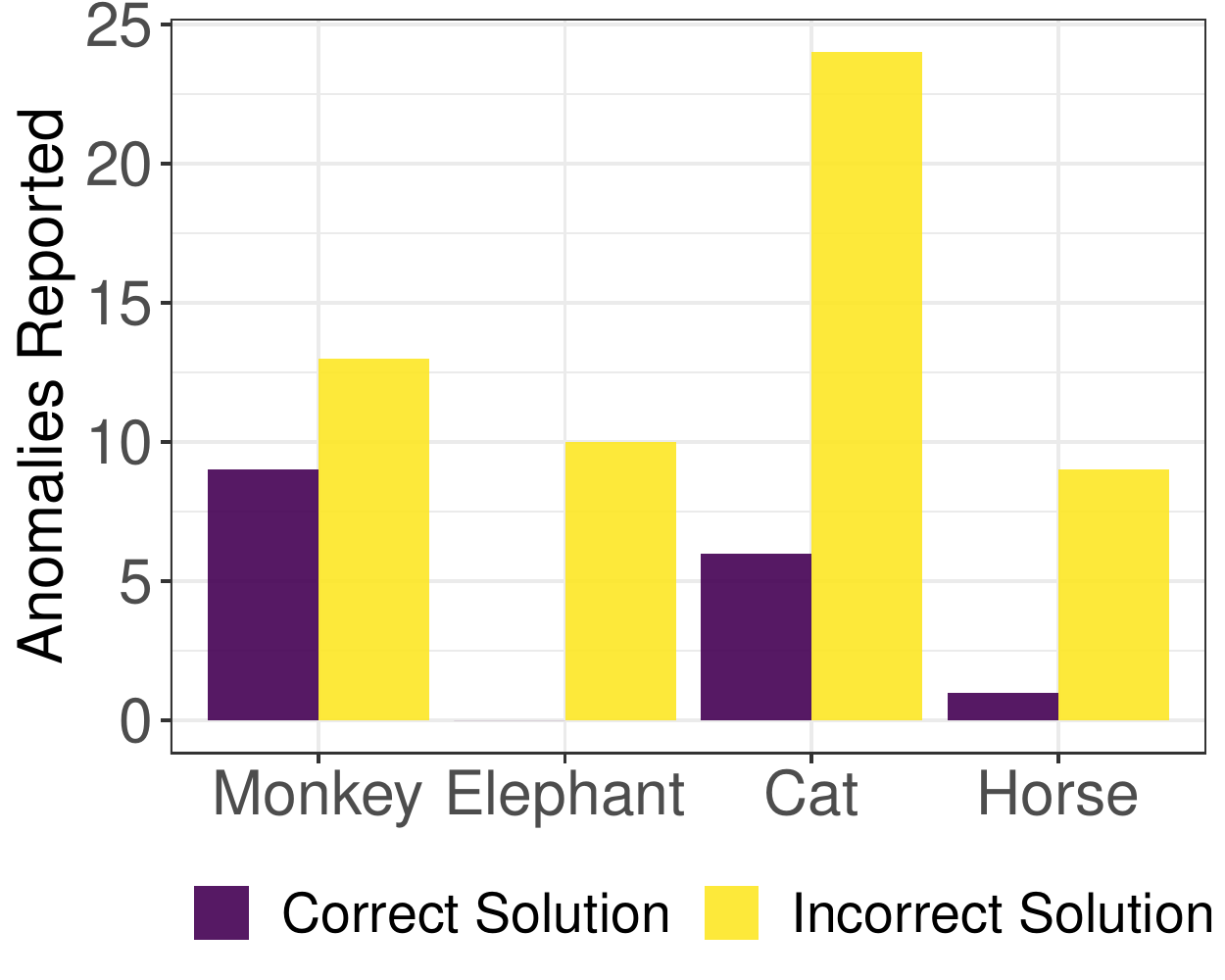}
      }
      \hfill
      \subfloat[\label{fig:correct-for-which-reported}Incorrect 
    solutions with anomalies.]{
      \centering
      \includegraphics[width=0.46\columnwidth]{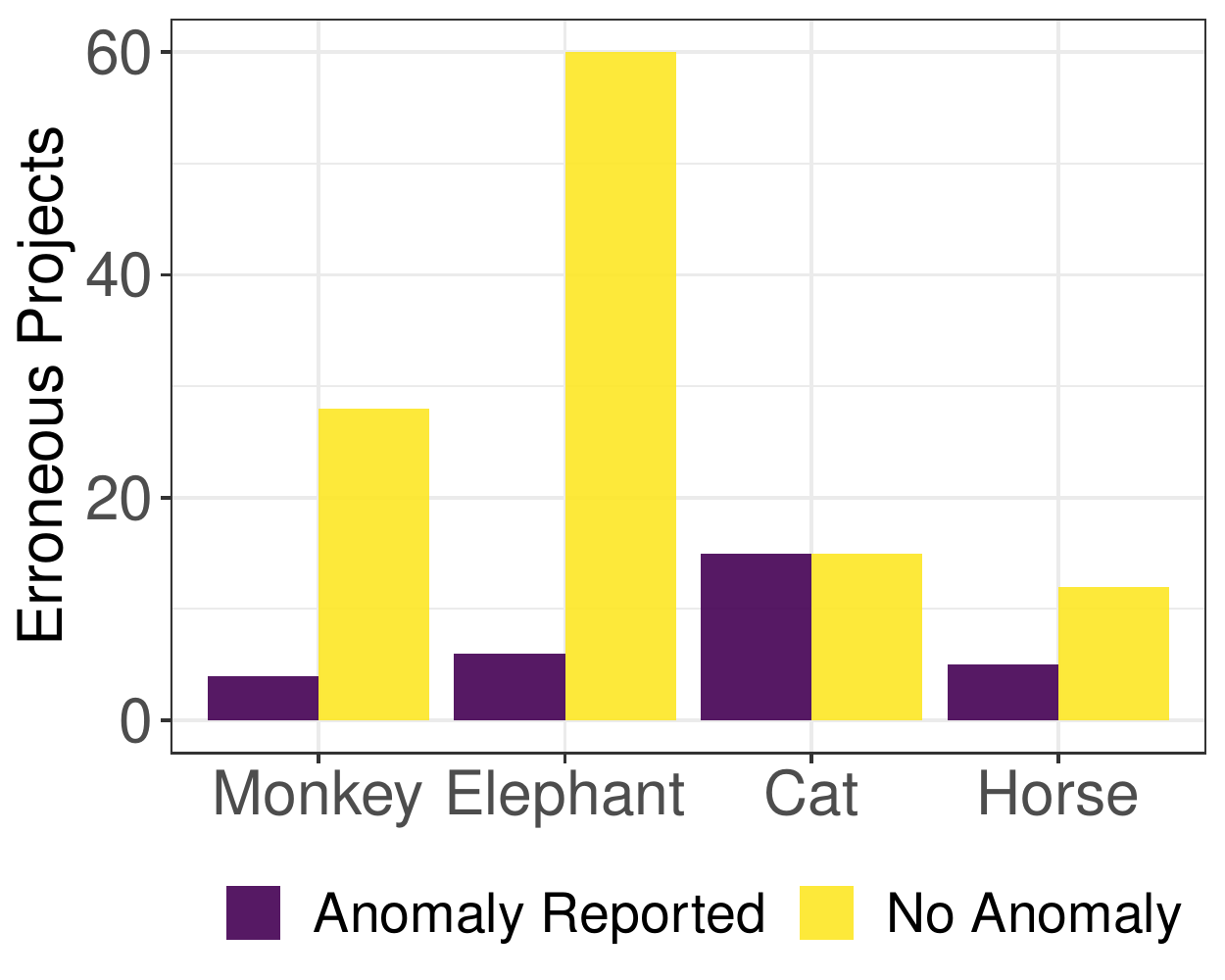}
    }
  \caption{\label{fig:reported}Relation of anomaly reports and correctness of solutions}
\vspace{-0.5em}
\end{figure}

\begin{figure}[t]
  \centering
  \includegraphics[width=4cm]{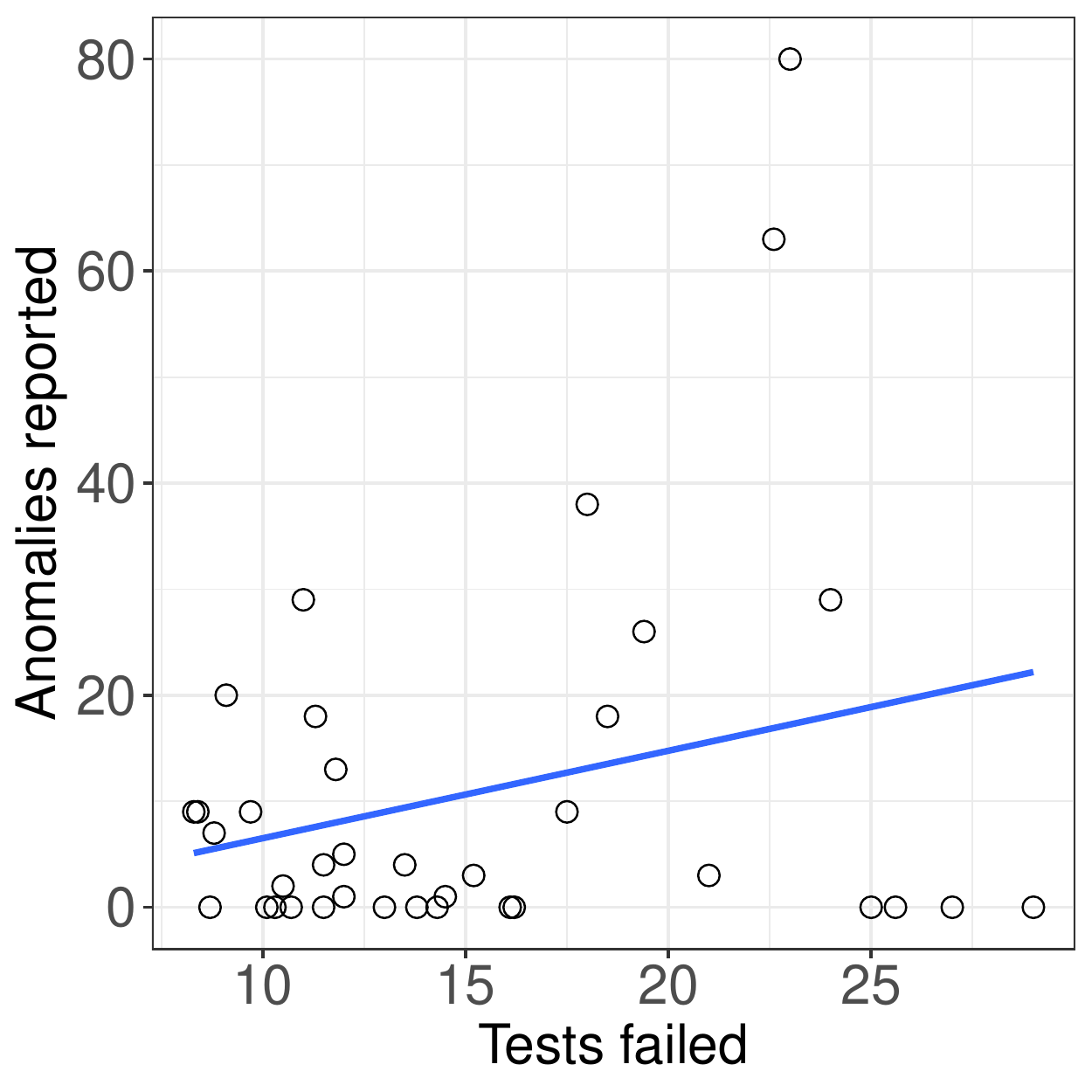}
  \caption{\label{fig:correct-fruit}Correlation between anomalies and failing tests}
\vspace{-0.5em}
\end{figure}

\begin{figure}[t]
  \centering
  \includegraphics[width=\columnwidth]{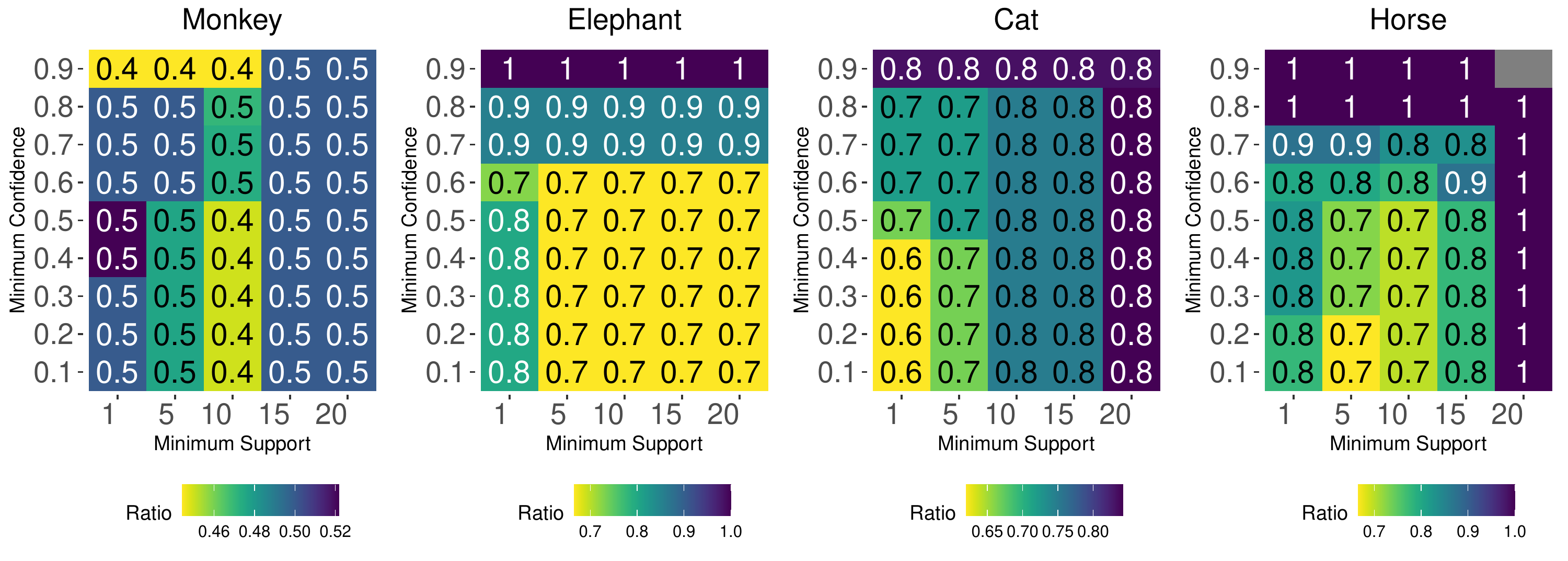}
  \caption{\label{fig:anomaly-ratio}Ratio of incorrect programs with anomalies reported vs. total number of programs with anomalies.}
\vspace{-0.5em}
\end{figure}

Fig.~\ref{fig:reported-on-correct} shows how many violations were
reported on correct/incorrect programs for the \monkey, \cat, \elephant,
and \horse tasks. Overwhelmingly, the programs for which anomalies were
reported are incorrect solutions. The proportion of correct programs
with anomalies is slightly higher for the \monkey program. This program
has two sprites (circus director and monkey), whereas only the
director is supposed to contain scripts. Manual inspection showed that
many anomalies are triggered by additional code in the monkey sprite,
which was not part of the specification (see Section~\ref{subsec:rq3}).
As correctness is a more fine-grained
question for the \fruit example, Fig.~\ref{fig:correct-fruit} shows the
correlation between anomalies reported and tests failed. There is a
weak correlation (Pearson correlation coefficient of~\num{0.27} with
$p = 0.105$), demonstrating that solutions with more errors
tend to have more anomalies reported, which supports the results on the other tasks.

Fig.~\ref{fig:correct-for-which-reported} shows how many of the
incorrect programs had anomalies reported.
While for the \cat example half the incorrect programs had an anomaly
reported, for the other tasks the proportion is lower. This is a
result of the number of patterns and anomalies mined with our
parameter settings, and lowering the confidence or minimum support
level would lead to more reported anomalies. 
However, lowering confidence and support thresholds may come at the price of more irrelevant anomalies: Fig.~\ref{fig:anomaly-ratio} shows the ratio of incorrect projects with anomalies reported to projects with anomalies reported in total for different parameter values. A higher ratio suggests a likely better quality of the reported anomalies, and Fig.~\ref{fig:anomaly-ratio} confirms that higher confidence and support increase the ratio.
%
%
%
\done{R1: Could there be heuristics, like for "What percent of students have to get a problem right, in order for the anomaly detector to spot bugs?"}
\done{R3: In the tuning of the parameters, the authors found that a very high threshold was necessary in most cases to keep from generating a huge number of anomalies. I wonder how the distribution of the type of anomaly changes as this threshold changes. That is, is it true that anomalies with higher confidence are in fact more likely to be “true bugs”?}

\done{R3: Figure 9 indicates a lot of erroneous programs that are missed. I wonder if these are caught with different parameter settings. From a teacher’s perspective, for example, the results for the \elephant program don’t seem all that usable: 10 anomalies were reported, all of which were erroneous programs, but an additional 60 programs were erroneous and not found by these settings.}
The number of missed incorrect solutions (Fig.~\ref{fig:correct-for-which-reported}) is
particularly notable for the \elephant example, where the overall
number of incorrect student solutions is also higher: For \elephant solutions to be considered correct, we required a \scratchblock{forever} loop with costume changes and \scratchblock{wait} blocks in between. Only~\num{25} student solutions used \scratchblock{forever} loops, while~\num{48} student solutions used \scratchblock{repeat times} loops instead, which we counted as incorrect. However, since this solution attempt is so common, it is unlikely to be reported as anomaly. In contrast, \num{11}~students used no loops at all, and thus their programs more likely result in anomalies. 
In general, if the dataset contains more diverse solutions, then fewer patterns will be found with high support.
This suggests that different use cases may require different parameter
settings. For example, formative assessment at intermediate points may require
different thresholds for support and confidence than summative assessment at the
end of the assignment.
%
Note that a further 46 solutions to the \elephant task are empty, i.e., consist only of the hat-block provided as starting point. Since at least two blocks are required in order to form a temporal property, no anomalies are reported for such empty projects (Fig.~\ref{fig:reported} only shows non-empty projects).



%

\summary{RQ2}{
There is a clear relationship between the number of anomalies 
and the correctness of a solution.
}

\subsection{RQ3: Which categories of anomalies can be identified?}
\label{subsec:rq3}%

\begin{figure}[t]
    \centering
    \includegraphics[width=\columnwidth]{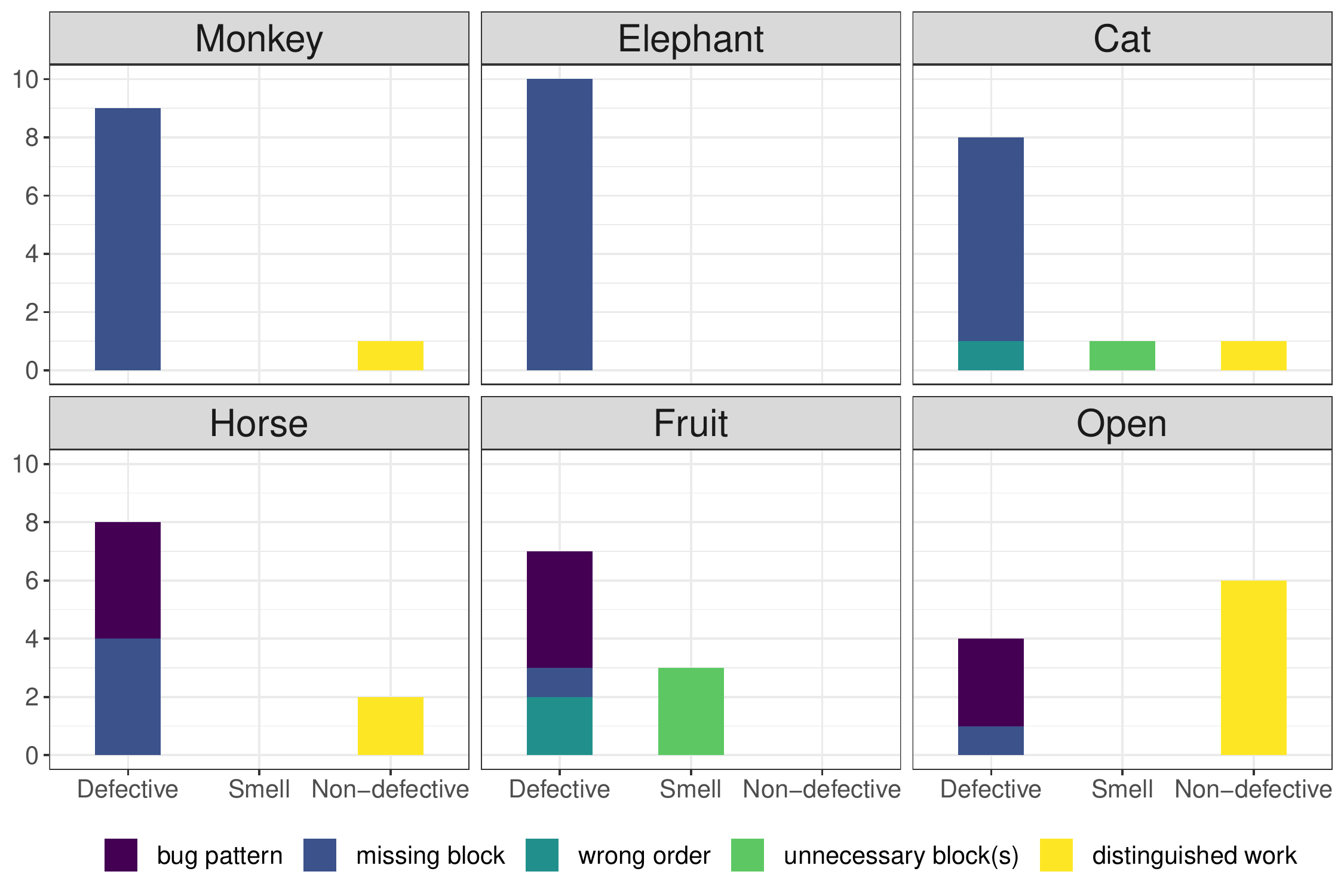}
	\ignore{R3: For the results in Figure 11, I wonder how many of these anomalies were, effectively, “the same”. For instance, for the elephant program, were all 10 erroneous programs missing the exact same block?}
    \caption{\label{fig:classification}Results of the manual anomaly classification}
\end{figure}

\newcommand{\analyzedvios}{\num{60}\xspace}
\newcommand{\defectivevios}{\num{46}\xspace}
\newcommand{\bugpatternvios}{\num{11}\xspace}
\newcommand{\missingblockvios}{\num{32}\xspace}
\newcommand{\wrongordervios}{\num{3}\xspace}
\newcommand{\projectspecificvios}{\num{35}\xspace}
\newcommand{\smellyviosuppercase}{\num{4}\xspace}
\newcommand{\smellyvioslowercase}{\num{4}\xspace}
\newcommand{\distinguishedvios}{\num{10}\xspace}

Fig.~\ref{fig:classification} summarizes the results of the manual
classification of the top ten anomalies for each of the datasets. In total,
\defectivevios of the \analyzedvios classified anomalies hint at defective
code, with \bugpatternvios anomalies in the subcategory bug patterns,
\missingblockvios in the subcategory missing blocks and \wrongordervios~in the
subcategory wrong order, \smellyviosuppercase~anomalies hinted at smelly scripts
with unnecessary code, and \distinguishedvios~hinted at non-defective,
distinguished work.

\begin{figure}[t]
    \centering
    \subfloat[Wrong block use to react to touches of
    the mouse pointer; similar in Fig.~\ref{fig:intro}.]{
    \centering
    \includegraphics[scale=0.15]{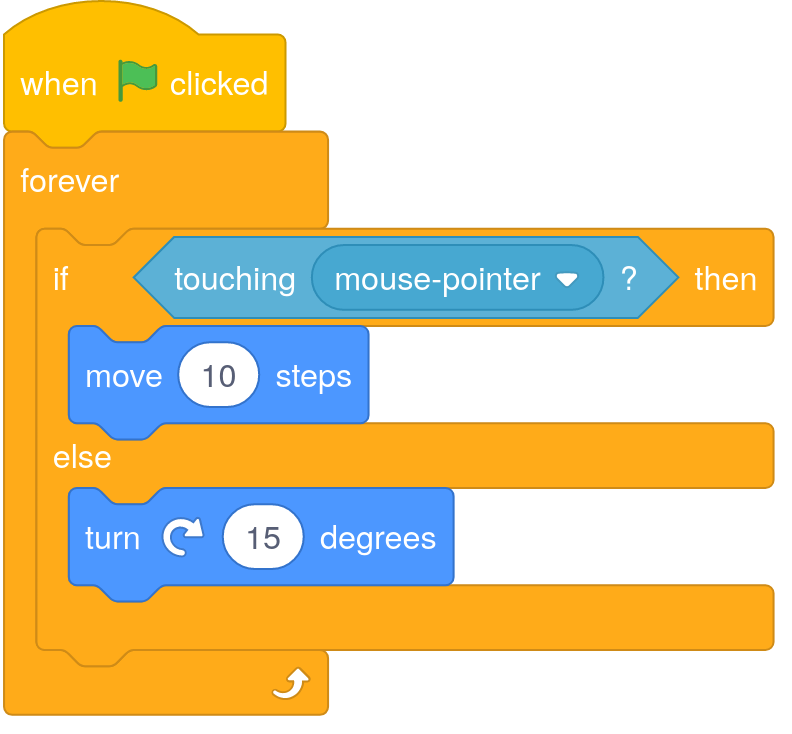}
    \label{fig:missing-block-script}
    }
    \hspace{3pt}
    \subfloat[The anomaly hints at the missing block in the script.]{
    \begin{tikzpicture}[node distance=4mm, scale=0.75, transform shape,
    cell/.style={rectangle,draw=black}, initial text={},
    space/.style={minimum height=1.2em,matrix of nodes,
    row sep=-\pgflinewidth,column sep=-\pgflinewidth}]
        \node[] (l0) {\scratchblock{when green flag}};
        \node[below left=of l0] (l1) {\scratchblock{forever}};
        \node[below=of l0] (l2) {\scratchblock{if-else}};
        \node[below=of l1] (l4) {\scratchblock{turn right}};
        \node[below right=of l4] (l3) {\mscratchblock{change graphic effect}};
        \path[trans] (l0) edge [bend right] node [pos=.5,label=right:{}] {} (l1);
        \path[trans] (l0) edge node [pos=.5,label=right:{}] {} (l2);
        \path[missing] (l0) edge [bend left] node [pos=.5,label=right:{}] {} (l3);
        \path[trans] (l0) edge [] node [pos=.5,label=right:{}] {} (l4);
        \path[trans] (l1) edge node [pos=.5,label=right:{}] {} (l2);
        \path[missing] (l1) edge [] node [pos=.5,label=right:{}] {} (l3);
        \path[trans] (l1) edge [bend right] node [pos=.5,label=right:{}] {} (l4);
        \path[trans] (l2) edge [loop right] node [pos=.5,label=right:{}] {} (l2);
        \path[trans] (l2) edge [bend left] node [pos=.5,label=right:{}] {} (l4);
        \path[missing] (l2) edge [bend right] node [pos=.5,label=right:{}] {} (l3);
        \path[missing] (l3) edge [loop below] node [pos=.5,label=right:{}] {} (l3);
        \path[missing] (l3) edge [] node [pos=.5,label=right:{}] {} (l2);
        \path[missing] (l3) edge [bend left] node [pos=.5,label=right:{}] {} (l4);
        \path[missing] (l4) edge node [pos=.5,label=right:{}] {} (l3);
        \path[trans] (l4) edge [] node [pos=.5,label=right:{}] {} (l2);
        \path[trans] (l4) edge [loop below] node [pos=.5,label=right:{}] {} (l4);
    \end{tikzpicture}
    }
    \caption{\label{fig:missing-block} The anomaly ranked third in the \horse
    task. It belongs to the missing block subcategory, as the block
    responsible for the required color change of the horse is missing.}
\vspace{-0.5em}
\end{figure}
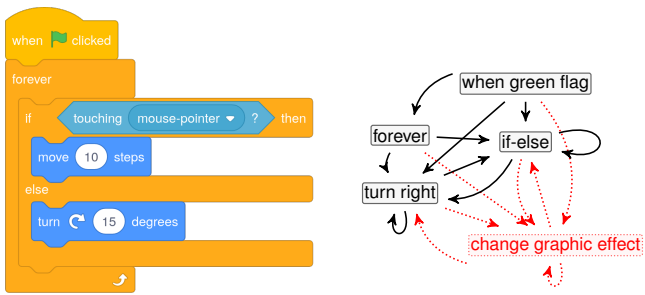

\begin{figure}[t]
    \centering
    \subfloat[The script makes the cat say something as soon as the game
    starts.]{
      \begin{minipage}[t]{0.46\columnwidth}
    \centering
    \includegraphics[scale=0.15]{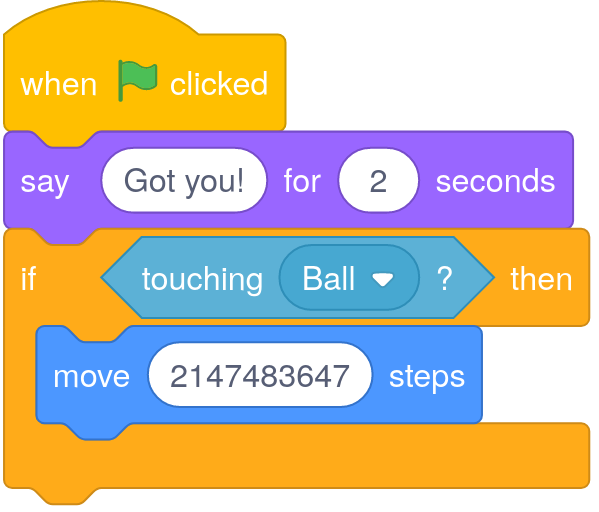}
    \end{minipage}
    \label{fig:wrong-order-script}
    }
    \hfill
    \subfloat[The anomaly shows that the say block should be used after the
    if block.]{
      \begin{minipage}[t]{0.46\columnwidth}
    \begin{tikzpicture}[node distance=6mm, scale=0.8, transform shape,
    cell/.style={rectangle,draw=black}, initial text={},
    space/.style={minimum height=1.2em,matrix of nodes,
    row sep=-\pgflinewidth,column sep=-\pgflinewidth}]
        \node[] (l0) {\scratchblock{when green flag}};
        \node[below left=of l0] (l1) {\scratchblock{if-then}};
        \node[below right=of l1] (l2) {\scratchblock{say for secs}};
        \path[trans] (l0) edge node [pos=.5,label=right:{}] {} (l1);
        \path[trans] (l0) edge node [pos=.5,label=right:{}] {} (l2);
        \path[missing] (l1) edge node [pos=.5,label=right:{}] {} (l2);
      \end{tikzpicture}
      \end{minipage}
    }
    \caption{\label{fig:wrong-order} The top ranked anomaly in the \cat
    task: It belongs to the wrong order subcategory, as the correct solution requires the cat to speak \textit{after}
    touching the ball.}
\vspace{-0.5em}
\end{figure}
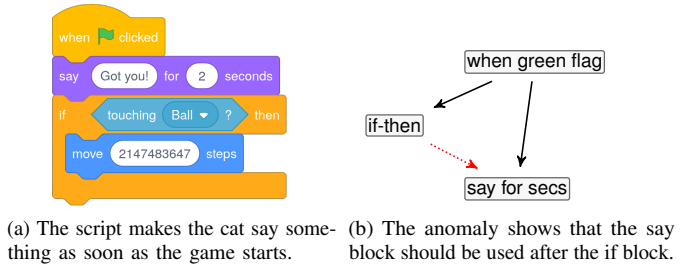

Except for the \open task, the majority of the detected anomalies hint at
defective code. Most defects~(\projectspecificvios out of \defectivevios) are
project-specific problems that a generic linter would miss: Missing blocks
and the wrong order of blocks. The anomalies in the tasks \cat, \elephant and
\monkey predominantly show that a specific block, which is essential for the
solution of the task, is missing in the student code.

As an example for the missing block category,
Fig.~\ref{fig:missing-block} shows a student solution for the \horse
task which does not have the block responsible for the required color
change. The anomaly shows the absence of this block, and therefore
provides important feedback for both the teacher and the student. The
student can be made aware of the missing block and the teacher can use
the anomaly as an opportunity to discuss in class when a task is
considered solved.

Fig.~\ref{fig:wrong-order} shows a project-specific anomaly of the wrong
order subcategory: Although the student's solution for the \cat task contains
most of the blocks necessary for solving the task, they are not in the correct
order---the script is defective. To help the student, a teacher can 
address the script flow in class or trace
the script step by step together with the student.

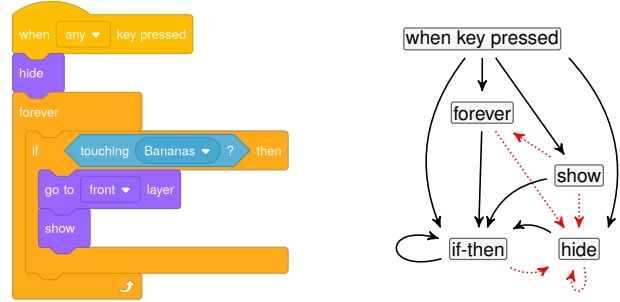
\begin{figure}[t]
    \centering
    \subfloat[The defective script first hides the sprite and then tries to show when touching another sprite. This does not work as sprites can only touch when visible.]{%
      \begin{minipage}[t]{0.52\columnwidth}
    \centering
    \includegraphics[scale=0.15]{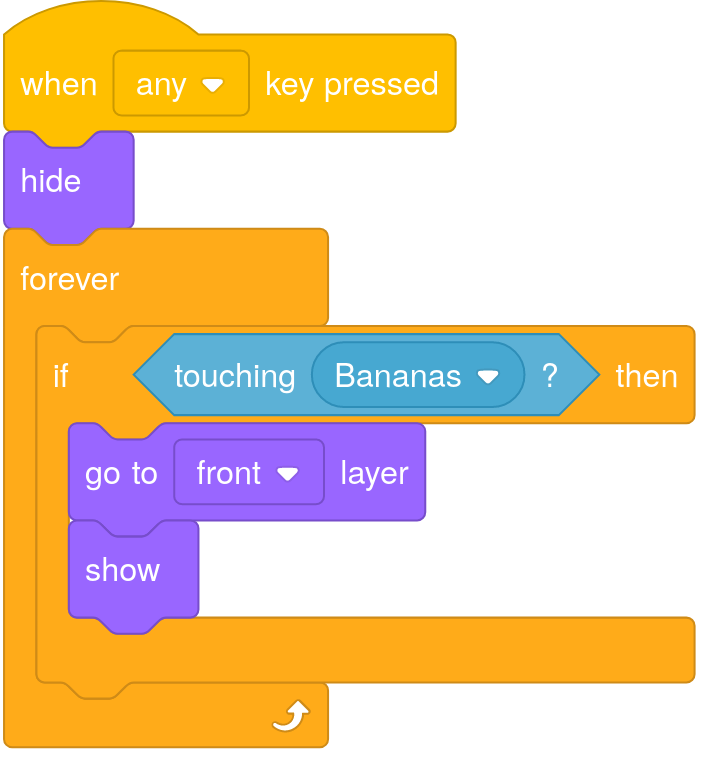}
    \end{minipage}
    \label{fig:bug-pattern-script}
    }
    \hfill
    \subfloat[The anomaly hints at the usual order of first showing the sprite before hiding it.]{
        \centering
    \begin{tikzpicture}[node distance=6mm, scale=0.8, transform shape,
    cell/.style={rectangle,draw=black}, initial text={},
    space/.style={minimum height=1.2em,matrix of nodes,
    row sep=-\pgflinewidth,column sep=-\pgflinewidth}]
        \node[] (l0) {\scratchblock{when key pressed}};
        \node[below=of l0] (l1) {\scratchblock{forever}};
        \node[below right=of l1] (l2) {\scratchblock{show}};
        \node[below=of l2] (l3) {\scratchblock{hide}};
        \node[left=of l3] (l4) {\scratchblock{if-then}};
        \path[trans] (l0) edge node [pos=.5,label=right:{}] {} (l2);
        \path[trans] (l0.south east) to [bend left] node [pos=.5, label=right:{}] {} (l3.north east);
        \path[trans] (l0) edge node [pos=.5,label=right:{}] {} (l1);
        \path[trans] (l0) edge [bend right] node [pos=.5,label=right:{}] {} (l4.north west);
        \path[missing] (l1) edge node [pos=.5,label=right:{}] {} (l3);
        \path[trans] (l1) edge node [pos=.5,label=right:{}] {} (l4);
        \path[trans] (l2) edge [bend right] node [pos=.5,label=right:{}] {} (l4);
        \path[missing] (l2) edge node [pos=.5,label=right:{}] {} (l1);
        \path[missing] (l2) edge node [pos=.5,label=right:{}] {} (l3);
        \path[trans] (l3) edge [bend right] node [pos=.5,label=right:{}] {} (l4);
        \path[missing] (l3) edge [loop below] node [pos=.5,label=right:{}] {} (l3);
        \path[missing] (l4) edge [bend right] node [pos=.5,label=right:{}] {} (l3);
        \path[trans] (l4) edge [loop left] node [pos=.5,label=right:{}] {} (l4);

      \end{tikzpicture}
    }
    \caption{\label{fig:bug-pattern} The anomaly ranked seventh in the \open
    task. It belongs to the bug pattern subcategory, as the ``Hide Before
    Touching'' defect could be detected by linters in principle.}
\vspace{-0.5em}
\end{figure}

Most of the anomalies in the subcategory bug pattern hinted at the bug
patterns~\cite{bugpatterns} Missing Loop Sensing (a condition that
should be checked repeatedly in a loop is checked only a single time),
Forever Inside Loop (an inner infinite loop prevents code in the outer
loop from being reached) and Terminated Loop (a loop is
unconditionally stopped after the first iteration) as implemented in
\litterbox. Fig.~\ref{fig:bug-pattern} shows an anomaly of the \open
task that shows a problem for which \litterbox does not yet define a
bug pattern, but which could be found by a generic checker: Before the
script checks if its sprite touches another sprite, the
\scratchblock{hide} block is executed. However, while being invisible
a sprite cannot touch other sprites, therefore the actions within the
true branch of the \scratchblock{if-then} are never executed. Based on
this anomaly, a teacher can not only help the individual student and
explain that the student should have used messages to coordinate the
program flow; the anomaly also provides clues about what
misunderstandings and misconceptions to touch upon in class.

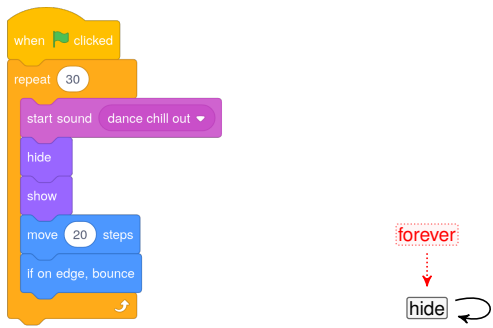
\begin{figure}[t]
    \centering
    \subfloat[The script implements a countdown using a timer variable and a
    conditional loop. As the value of the timer will always be zero, the
    conditional after the loop is unnecessary.]{
      \begin{minipage}[t]{0.46\columnwidth}
    \centering
    \includegraphics[scale=0.15]{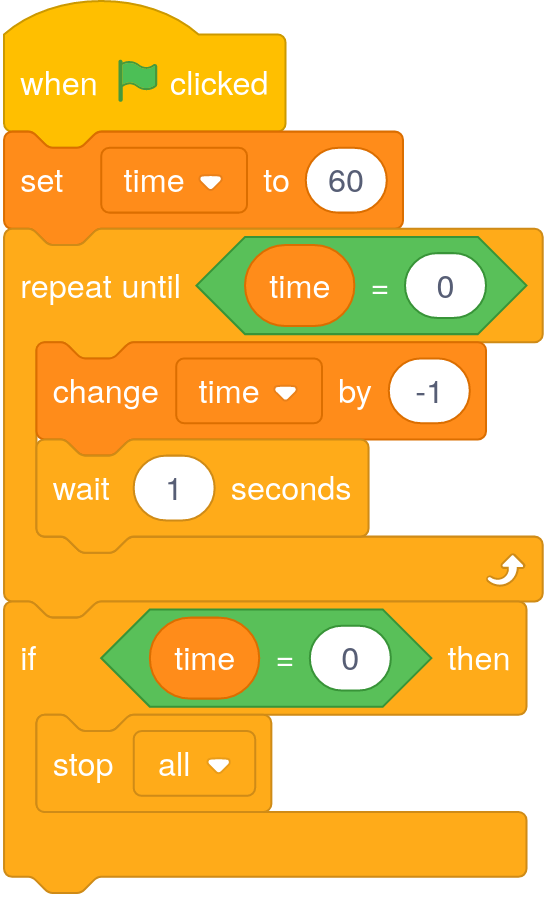}
    \end{minipage}
    \label{fig:unneccessary-script}
    }
    \hfill
    \subfloat[Even though the anomaly does not directly mark the conditional as
    unnecessary, it directs the attention to the unusual usage and therefore
    the smelly part of the script.]{
      \begin{minipage}[t]{0.44\columnwidth}
        \centering
    \begin{tikzpicture}[node distance=6mm, scale=0.8, transform shape,
    cell/.style={rectangle,draw=black}, initial text={},
    space/.style={minimum height=1.2em,matrix of nodes,
    row sep=-\pgflinewidth,column sep=-\pgflinewidth}]
        \node[] (l0) {\scratchblock{if-then}};
        \node[below=of l0] (l1) {\scratchblock{change variable by}};
        \path[trans] (l1) edge [bend right] node [pos=.5,label=right:{}] {} (l0);
        \path[missing] (l0) edge [bend right] node [pos=.5,label=right:{}] {} (l1);
        \path[missing] (l0) edge [loop right] node [pos=.5,label=right:{}] {} (l0);
      \end{tikzpicture}
      \end{minipage}
    \label{fig:unneccessary-violation}
    }
    \caption{\label{fig:unneccessary} The anomaly ranked second in
    the \fruit task. It belongs to the unnecessary block(s) subcategory, as the
    anomaly hints at a script that contains redundant blocks.}
\vspace{-0.5em}
\end{figure}

Besides anomalies that indicate defective code, there
are~\smellyvioslowercase~cases of smelly code with
extraneous scripts or blocks that do not influence the program
behavior, but negatively affect the code quality. In the script
in Fig.~\ref{fig:unneccessary-script}, the student programmed a
countdown using a timer variable and a conditional loop breaking when
the timer is equal to zero. Subsequently, the script uses an
\scratchblock{if-then} block to check if the timer equals zero. This
block is redundant, since the conditional loop already determines the
countdown to stop as soon as the timer is set to zero. Even if the
anomaly in Fig.~\ref{fig:unneccessary-violation} does not explicitly
state that the \scratchblock{if-then} is redundant, it directs the
attention to the conditional. Building on such examples of smelly
code, broader concepts from software engineering, such as code quality
issues, can be incorporated into teaching.

\begin{figure}[t]
    \centering
    \subfloat[The script implements an animation and works fine.]{
    \centering
    \includegraphics[scale=0.15]{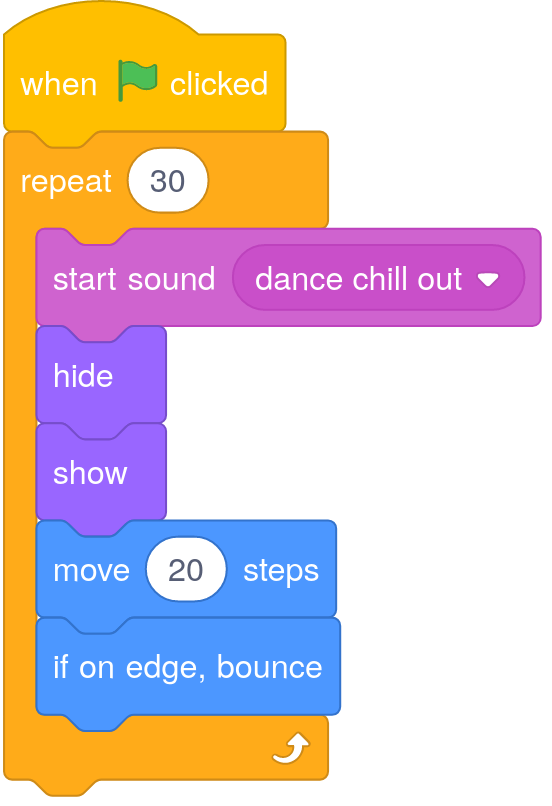}
    \label{fig:distinguished-script}
    }
    %
    \hfill
    \subfloat[The anomaly suggests to adhere to a pattern similar to the pattern violated in Fig.~\ref{fig:bug-pattern}]{
      \begin{minipage}[t]{0.6\columnwidth}
        \centering
    \begin{tikzpicture}[node distance=6mm, scale=0.8, transform shape,
    cell/.style={rectangle,draw=black}, initial text={},
    space/.style={minimum height=1.2em,matrix of nodes,
    row sep=-\pgflinewidth,column sep=-\pgflinewidth}]
        \node[] (l0) {\mscratchblock{forever}};
        \node[below=of l0] (l1) {\scratchblock{hide}};
        \path[trans] (l1) edge [loop right] node [pos=.5,label=right:{}] {} (l1);
        \path[missing] (l0) edge node [pos=.5,label=right:{}] {} (l1);
      \end{tikzpicture}
      \end{minipage}
    }
    \caption{\label{fig:distinguished} The anomaly ranked ninth in the \open
    task belongs to the distinguished work subcategory, as satisfying
    the pattern would contribute to neither quality nor correctness.}
\vspace{-0.5em}
\end{figure}

There are \distinguishedvios cases where there are scripts that trigger
anomalies, even though the underlying code is not erroneous. The majority of these are,
unsurprisingly, in the \open task, where students were free to implement games
of their choice, based on common previous tasks. For example, although the
anomaly in Fig.~\ref{fig:distinguished} suggests to use a
\scratchblock{forever} loop, the programmed animation is neither defective
nor smelly and therefore does not need to be changed. In the context of the
closed tasks, the anomalies either indicated that the students programmed
something different from or additional to what was required in the task, for example, added code to
sprites which usually are left empty. Both types of anomalies can be helpful for the classroom
context since the student can be made aware of the actual task and the
teacher can (if necessary) adjust teaching activities and pacing, or acknowledge and reward creative extensions of the tasks to encourage student creativity.

\summary{RQ3}{Out of \analyzedvios classified anomalies, \defectivevios
hinted at defective code, \smellyvioslowercase hinted at smelly code and
\distinguishedvios hinted at distinguished student work. All of these
anomalies provide valuable feedback for teachers.}

\section{Related Work}

Alternative approaches for analyzing \scratch programs introduced in
Section~\ref{subsec:background-scratch} include linting, testing, and
verification. All of these require some sort of
prior, manual work---
tests, checks or specifications,
whereas anomaly detection requires no manual work. Furthermore, in
contrast to generic linters anomaly detection can also find
project-specific bugs; on the other hand, anomalies may help to
identify new, previously unknown generic checks to implement in
linters, such as the hide-show defect (Fig.~\ref{fig:bug-pattern})
we discovered in our analysis.
The quality and number of reported anomalies, however, depends on the
underlying dataset, 
 the number of students, and their overall
progress in the programming assignment; these are factors we plan to
study in our future work.

Our approach for anomaly detection in \scratch is based on the \jadet
tool, which is originally designed to analyze object usage models
for \java{} objects~\cite{jadet}. We chose this approach
because approaches using the version history~\cite{dynamine} are not
applicable on \scratch, and our motivation from an educational point
of view is to find anomalies in the temporal relation of blocks,
rather than relations between variables and method
calls~\cite{li2005pr} or patterns of interactions of multiple
objects~\cite{grouminer}. However, many different anomaly mining
techniques for software have been proposed over the years, and others
may also be applicable to our specific domain.


\section{Conclusions}

With programming education becoming more prevalent, even at earlier
ages, there is an increasing demand for tools to support educators and
learners. To the best of our knowledge, this paper is the first
proposal to use anomaly detection on \scratch student code. Anomaly
detection 
requires no manual
specification effort, and, as our evaluation demonstrated, is
nevertheless effective at finding relevant issues.

Our initial investigation achieved promising results, but
also raised many interesting follow-up questions for future
investigation:
The specific technique of anomaly detection we implemented has several
parameters, 
and other anomaly detection techniques might be able to find other or more interesting anomalies. Understanding what techniques and parameters lead to the results that are most helpful will require further experiments, and a better understanding of when and how teachers and learners would apply
 anomaly detection. 
A related question is how to best present anomalies to teachers and
students in a way that helps them to understand the problem with their
code, and how to fix it. Often, the pattern violated by an anomaly may
be able to serve as a hint on a correction.

%

While programming and code quality are essential aspects of software
engineering education, 
 anomaly detection is 
 applicable to any
software engineering artifacts for which patterns can be formalized. It may therefore be possible to support education with respect
to all phases of the software engineering life cycle.

\done{R1: Could the anomaly detection tool employed in this experiment also be used by teachers to help isolate student errors in real time?  (Or perhaps by students on their own?)  E.g., when a group of students is working on a problem for which there already is a history of answers.  Perhaps they could conclude by forecasting such situations, in preparation for their next steps?}

\done{R1: How much of the foundations of this analysis would teachers have to understand, to use this methodology?  How deeply do they need to understand its strengths and limitations?}

\done{R1: In the visually oriented SCRATCH, one could argue that a teacher can determine the correct running of a program just by watching it run.  So, the value of anomaly detection lies in analyzing and isolating the bugs, I think.  Which makes the useful role of anomaly detection slightly harder to demonstrate.}

\todo[inline]{R1: Is it likely that a typical SCRATCH teacher could only analyze student-written programs using these tools, if the programs were pre-existing ones which a more seasoned person had run all the pattern detection software on?}

\done{R3: Students often make mistakes in the very same ways. (Often this may point to a confusing statement in the problem or even in the programming language.) Nonetheless, these may be incorrect programs that are not “anomalies”. I wonder how frequent these are and what mitigation strategies there are in this approach. The fact that existing research shows that SCRATCH programmers tend to learn some specific bad habits would seem to bolster this concern. In the paper’s proposed approach, this contributes to the contribution of the confidence associated with a pattern violation. Those violations that occur in more samples have lower confidence as being true anomalies.}

\section*{Acknowledgements}
This work is supported by DFG project FR 2955/3-1 ``Testing,
Debugging, and Repairing Blocks-based Programs''.

\balance

\bibliographystyle{IEEEtranS}
\bibliography{references}

\begin{thebibliography}{10}
\providecommand{\url}[1]{#1}
\csname url@samestyle\endcsname
\providecommand{\newblock}{\relax}
\providecommand{\bibinfo}[2]{#2}
\providecommand{\BIBentrySTDinterwordspacing}{\spaceskip=0pt\relax}
\providecommand{\BIBentryALTinterwordstretchfactor}{4}
\providecommand{\BIBentryALTinterwordspacing}{\spaceskip=\fontdimen2\font plus
\BIBentryALTinterwordstretchfactor\fontdimen3\font minus
  \fontdimen4\font\relax}
\providecommand{\BIBforeignlanguage}[2]{{%
\expandafter\ifx\csname l@#1\endcsname\relax
\typeout{** WARNING: IEEEtranS.bst: No hyphenation pattern has been}%
\typeout{** loaded for the language `#1'. Using the pattern for}%
\typeout{** the default language instead.}%
\else
\language=\csname l@#1\endcsname
\fi
#2}}
\providecommand{\BIBdecl}{\relax}
\BIBdecl

\bibitem{aivaloglou2016kids}
E.~Aivaloglou and F.~Hermans, ``{How kids code and how we know: An exploratory
  study on the Scratch repository},'' in \emph{ACM Conference on International
  Computing Education Research}.\hskip 1em plus 0.5em minus 0.4em\relax ACM,
  2016, pp. 53--61.

\bibitem{ala2005survey}
K.~M. Ala-Mutka, ``A survey of automated assessment approaches for programming
  assignments,'' \emph{Computer science education}, vol.~15, no.~2, pp.
  83--102, 2005.

\bibitem{beller2017developer}
M.~Beller, G.~Gousios, A.~Panichella, S.~Proksch, S.~Amann, and A.~Zaidman,
  ``{Developer testing in the ide: Patterns, beliefs, and behavior},''
  \emph{IEEE Transactions on Software Engineering}, vol.~45, no.~3, pp.
  261--284, 2017.

\bibitem{blondeau2017testing}
V.~Blondeau, A.~Etien, N.~Anquetil, S.~Cresson, P.~Croisy, and S.~Ducasse,
  ``{What are the testing habits of developers? A case study in a large IT
  company},'' in \emph{2017 IEEE International Conference on Software
  Maintenance and Evolution (ICSME)}.\hskip 1em plus 0.5em minus 0.4em\relax
  IEEE, 2017, pp. 58--68.

\bibitem{boe2013hairball}
B.~Boe, C.~Hill, M.~Len, G.~Dreschler, P.~Conrad, and D.~Franklin, ``{Hairball:
  Lint-inspired static analysis of scratch projects},'' in \emph{ACM Technical
  Symposium on Computer Science Education}.\hskip 1em plus 0.5em minus
  0.4em\relax ACM, 2013, pp. 215--220.

\bibitem{chang2018scratch}
Z.~Chang, Y.~Sun, T.-Y. Wu, and M.~Guizani, ``{Scratch analysis Tool (SAT): a
  modern scratch project analysis tool based on ANTLR to assess computational
  thinking skills},'' in \emph{2018 14th International Wireless Communications
  \& Mobile Computing Conference (IWCMC)}.\hskip 1em plus 0.5em minus
  0.4em\relax IEEE, 2018, pp. 950--955.

\bibitem{douce2005automatic}
C.~Douce, D.~Livingstone, and J.~Orwell, ``{Automatic test-based assessment of
  programming: A review},'' \emph{Journal on Educational Resources in Computing
  (JERIC)}, vol.~5, no.~3, p.~4, 2005.

\bibitem{edwards2019can}
S.~Edwards, J.~Spacco, and D.~Hovemeyer, ``{Can Industrial-Strength Static
  Analysis Be Used to Help Students Who Are Struggling to Complete Programming
  Activities?}'' in \emph{Proceedings of the 52nd Hawaii International
  Conference on System Sciences}, 2019.

\bibitem{eisenbarth}
T.~Eisenbarth, R.~Koschke, and G.~Vogel, ``Static object trace extraction for
  programs with pointers,'' \emph{Journal of Systems and Software}, vol.~77,
  no.~3, pp. 263--284, 2005.

\bibitem{bugpatterns}
C.~Frädrich, F.~Obermüller, N.~Körber, U.~Heuer, and G.~Fraser, ``{Common
  Bugs in Scratch Programs},'' in \emph{Proceedings of the 2020 ACM Conference
  on Innovation and Technology in Computer Science Education}, ser. ITiCSE
  '20.\hskip 1em plus 0.5em minus 0.4em\relax ACM, 2020.

\bibitem{katharina}
K.~Geldreich, A.~Funke, and P.~Hubwieser, ``A programming circus for primary
  schools,'' in \emph{ISSEP 2016}, 2016, pp. 49--50.

\bibitem{Grover.2017}
S.~Grover, ``{Assessing Algorithmic and Computational Thinking in K-12: Lessons
  from a Middle School Classroom},'' in \emph{Emerging Research, Practice, and
  Policy on Computational Thinking}, P.~J. Rich and C.~B. Hodges, Eds.\hskip
  1em plus 0.5em minus 0.4em\relax Cham: {Springer International Publishing},
  2017, pp. 269--288.

\bibitem{Grover.2020}
S.~Grover, V.~Sedgwick, and K.~Powers.

\bibitem{gulwani2016automated}
S.~Gulwani, I.~Radicek, and F.~Zuleger, ``{Automated Clustering and Program
  Repair for Introductory Programming Assignments},'' \emph{arXiv preprint
  arXiv:1603.03165}, 2016.

\bibitem{Hassinen.2006}
M.~Hassinen and H.~M{\"a}yr{\"a}, ``{Learning Programming by Programming: a
  Case Study},'' in \emph{Proceedings KolliCalling}, A.~Berglund and
  M.~Wigbberg, Eds., 2006, pp. 117--119.

\bibitem{hermans2016code}
F.~Hermans and E.~Aivaloglou, ``{Do code smells hamper novice programming? A
  controlled experiment on Scratch programs},'' in \emph{Int. Conference on
  Program Comprehension}.\hskip 1em plus 0.5em minus 0.4em\relax IEEE, 2016,
  pp. 1--10.

\bibitem{hermans2016a}
F.~Hermans, K.~T. Stolee, and D.~Hoepelman, ``{Smells in Block-Based
  Programming Languages},'' in \emph{2016 {{IEEE Symposium}} on {{Visual
  Languages}} and {{Human}} - {{Centric Computing}} ( {{VL}} / {{HCC}}
  )}.\hskip 1em plus 0.5em minus 0.4em\relax {IEEE}, 2016, pp. 68--72.

\bibitem{AutomataTheory}
J.~E. Hopcroft, R.~Motwani, and J.~D. Ullman, \emph{{Introduction to automata
  theory, languages, and computation, 3rd Edition}}, ser. Pearson international
  edition.\hskip 1em plus 0.5em minus 0.4em\relax Addison-Wesley, 2007.

\bibitem{hovemeyer2004finding}
D.~Hovemeyer and W.~Pugh, ``{Finding bugs is easy},'' \emph{ACM SIGPLAN
  Notices}, vol.~39, no.~12, pp. 92--106, 2004.

\bibitem{Hubwieser.2013}
P.~Hubwieser, J.~Magenheim, A.~M{\"u}hling, and A.~Ruf, ``{Towards a
  conceptualization of pedagogical content knowledge for computer science},''
  in \emph{ICER '13}, B.~Simon, A.~Clear, and Q.~Cutts, Eds.\hskip 1em plus
  0.5em minus 0.4em\relax ACM, 2013, p.~1.

\bibitem{ihantola2010review}
P.~Ihantola, T.~Ahoniemi, V.~Karavirta, and O.~Sepp{\"a}l{\"a}, ``Review of
  recent systems for automatic assessment of programming assignments,'' in
  \emph{Koli Calling International Conference on Computing Education
  Research}.\hskip 1em plus 0.5em minus 0.4em\relax ACM, 2010, pp. 86--93.

\bibitem{Insa.2016}
D.~Insa and J.~Silva, ``{Semi-Automatic Assessment of Unrestrained Java
  Code},'' in \emph{Proceedings of the 2015 ACM Conference on Innovation and
  Technology in Computer Science Education}, ser. ITiCSE '15, V.~Dagien~{\.
  {e}}, C.~Schulte, and T.~Jevsikova, Eds.\hskip 1em plus 0.5em minus
  0.4em\relax ACM, 2015, pp. 39--44.

\bibitem{johnson2016itch}
D.~E. Johnson, ``{ITCH: Individual Testing of Computer Homework for Scratch
  Assignments},'' in \emph{Proceedings of the 47th ACM Technical Symposium on
  Computing Science Education}.\hskip 1em plus 0.5em minus 0.4em\relax ACM,
  2016, pp. 223--227.

\bibitem{krusche2018artemis}
S.~Krusche and A.~Seitz, ``{ArTEMiS: An automatic assessment management system
  for interactive learning},'' in \emph{Proceedings of the 49th ACM Technical
  Symposium on Computer Science Education}, 2018, pp. 284--289.

\bibitem{li2005pr}
Z.~Li and Y.~Zhou, ``{PR-Miner: automatically extracting implicit programming
  rules and detecting violations in large software code},'' \emph{ACM SIGSOFT
  Software Engineering Notes}, vol.~30, no.~5, pp. 306--315, 2005.

\bibitem{dynamine}
B.~Livshits and T.~Zimmermann, ``{Dynamine: finding common error patterns by
  mining software revision histories},'' \emph{ACM SIGSOFT Software Engineering
  Notes}, vol.~30, no.~5, pp. 296--305, 2005.

\bibitem{Magnusson.1999}
S.~Magnusson, J.~Krajcik, and H.~Borko, ``{Nature, Sources, and Development of
  Pedagogical Content Knowledge for Science Teaching},'' in \emph{Examining
  Pedagogical Content Knowledge}, ser. Science {\&} Technology Education
  Library, J.~Gess-Newsome and N.~G. Lederman, Eds.\hskip 1em plus 0.5em minus
  0.4em\relax Dordrecht: {Kluwer Academic Publishers}, 2002, vol.~6, pp.
  95--132.

\bibitem{maloney2010}
J.~Maloney, M.~Resnick, N.~Rusk, B.~Silverman, and E.~Eastmond, ``{The Scratch
  Programming Language and Environment},'' \emph{ACM Transactions on Computing
  Education (TOCE)}, vol.~10, p.~16, 11 2010.

\bibitem{Mayring.2015}
P.~Mayring, ``{Qualitative content analysis: Theoretical background and
  procedures},'' in \emph{Approaches to qualitative research in mathematics
  education}.\hskip 1em plus 0.5em minus 0.4em\relax Springer, 2015, pp.
  365--380.

\bibitem{meerbaum2011habits}
O.~Meerbaum-Salant, M.~Armoni, and M.~Ben-Ari, ``Habits of programming in
  scratch,'' in \emph{16th Annual Joint Conference on Innovation and Technology
  in Computer Science Education}.\hskip 1em plus 0.5em minus 0.4em\relax ACM,
  2011, pp. 168--172.

\bibitem{grouminer}
T.~T. Nguyen, H.~A. Nguyen, N.~H. Pham, J.~M. Al-Kofahi, and T.~N. Nguyen,
  ``{Graph-based mining of multiple object usage patterns},'' in
  \emph{Proceedings of the 7th joint meeting of the European Software
  Engineering Conference and the ACM SIGSOFT symposium on the Foundations of
  Software Engineering}, 2009, pp. 383--392.

\bibitem{pecorelli2020testing}
F.~Pecorelli, G.~Catolino, F.~Ferrucci, A.~De~Lucia, and F.~Palomba, ``{Testing
  of Mobile Applications in the Wild: A Large-Scale Empirical Study on Android
  Apps},'' in \emph{Proceedings of the 28th International Conference on Program
  Comprehension}, 2020, pp. 296--307.

\bibitem{perelman2014test}
D.~Perelman, S.~Gulwani, and D.~Grossman, ``Test-driven synthesis for automated
  feedback for introductory computer science assignments,'' \emph{Data Mining
  for Educational Assessment and Feedback (ASSESS 2014)}, 2014.

\bibitem{pham2014enablers}
R.~Pham, S.~Kiesling, O.~Liskin, L.~Singer, and K.~Schneider, ``{Enablers,
  inhibitors, and perceptions of testing in novice software teams},'' in
  \emph{Proceedings of the 22nd ACM SIGSOFT International Symposium on
  Foundations of Software Engineering}, 2014, pp. 30--40.

\bibitem{Rahimi.2017}
E.~Rahimi, E.~Barendsen, and I.~Henze, ``{Identifying Students' Misconceptions
  on Basic Algorithmic Concepts Through Flowchart Analysis},'' in
  \emph{Informatics in Schools}, V.~Dagien~{\. {e}} and A.~Hellas, Eds.\hskip
  1em plus 0.5em minus 0.4em\relax Cham: {Springer International Publishing},
  2017, vol. 10696, pp. 155--168.

\bibitem{robles2017software}
G.~Robles, J.~Moreno-Le{\'o}n, E.~Aivaloglou, and F.~Hermans, ``{Software
  clones in scratch projects: On the presence of copy-and-paste in
  computational thinking learning},'' in \emph{2017 IEEE 11th International
  Workshop on Software Clones (IWSC)}.\hskip 1em plus 0.5em minus 0.4em\relax
  IEEE, 2017, pp. 1--7.

\bibitem{Sentance.2017}
S.~Sentance and A.~Csizmadia, ``{Computing in the curriculum: Challenges and
  strategies from a teacher's perspective},'' \emph{Education and Information
  Technologies}, vol.~22, no.~2, pp. 469--495, 2017.

\bibitem{VerifiedFromScratch}
A.~Stahlbauer, C.~Frädrich, and G.~Fraser, ``{Verified from Scratch: Program
  Analysis for Learners’ Programs},'' in \emph{{ASE}}.\hskip 1em plus 0.5em
  minus 0.4em\relax {IEEE}, 2020.

\bibitem{whisker}
A.~Stahlbauer, M.~Kreis, and G.~Fraser, ``Testing scratch programs
  automatically,'' in \emph{Proceedings of the 2019 27th ACM Joint Meeting on
  European Software Engineering Conference and Symposium on the Foundations of
  Software Engineering}, 2019, pp. 165--175.

\bibitem{techapaloku2017b}
P.~Techapalokul and E.~Tilevich, ``{Quality Hound — An online code smell
  analyzer for scratch programs},'' in \emph{2017 IEEE Symposium on Visual
  Languages and Human-Centric Computing (VL/HCC)}, Oct 2017, pp. 337--338.

\bibitem{wang2016bugram}
S.~Wang, D.~Chollak, D.~Movshovitz-Attias, and L.~Tan, ``{Bugram: bug detection
  with n-gram language models},'' in \emph{Proceedings of the 31st IEEE/ACM
  International Conference on Automated Software Engineering}, 2016, pp.
  708--719.

\bibitem{diss}
A.~Wasylkowski, ``{Object Usage: Patterns and Anomalies},'' Ph.D. dissertation,
  Saarland University, 2010.

\bibitem{wasylkowski2011mining}
A.~Wasylkowski and A.~Zeller, ``{Mining temporal specifications from object
  usage},'' \emph{Automated Software Engineering}, vol.~18, no. 3-4, pp.
  263--292, 2011.

\bibitem{jadet}
A.~Wasylkowski, A.~Zeller, and C.~Lindig, ``{Detecting Object Usage
  Anomalies},'' in \emph{{Proceedings of the the 6th Joint Meeting of the
  European Software Engineering Conference and the ACM SIGSOFT Symposium on The
  Foundations of Software Engineering}}.\hskip 1em plus 0.5em minus 0.4em\relax
  ACM, 2007, pp. 35--44.

\end{thebibliography}
 
\end{document}